\documentclass[hyper,letterpaper,11pt]{JHEP3}
\usepackage{amsmath,amssymb,multirow,array}
\usepackage{cite}
\newcommand{\ddT}[1]{\frac{\partial #1}{\partial T}}
\newcommand{\ddm}[1]{\frac{\partial #1}{\partial\mu}}

\newcommand{\Sone}{\mathbb{S}^1}

\newcommand{\hs}{*}

\newcommand{\prn}[1]{\left ( #1 \right )}
\newcommand{\brk}[1]{\left [ #1 \right ]}

\title{Thermodynamics, gravitational anomalies and cones}

\author{Kristan Jensen$^a$, R. Loganayagam$^b$, and Amos Yarom$^{c}$ \\
$^a$ Department of Physics and Astronomy, University of Victoria, Victoria, BC V8W 3P6, Canada \\
$^b$ Junior Fellow, Harvard Society of Fellows, Harvard University, Cambridge, MA 02138. \\
$^c$ Department of Physics, Technion, Haifa 32000, Israel \\
Email: kristanj@uvic.ca, nayagam@physics.harvard.edu, ayarom@physics.technion.ac.il
}

\abstract{
By studying the Euclidean partition function on a cone, we argue that pure and mixed gravitational anomalies generate a ``Casimir momentum'' which manifests itself as parity violating coefficients in the hydrodynamic stress tensor and charge current. The coefficients generated by these anomalies enter at a lower order in the hydrodynamic gradient expansion than would be naively expected. In $1+1$ dimensions, the gravitational anomaly affects coefficients at zeroth order in the gradient expansion. The mixed anomaly in $3+1$ dimensions controls the value of coefficients at first order in the gradient expansion.
}

\begin{document}

\section{Introduction}
\label{S:intro}
Anomalies constitute a set of  fascinating phenomena in field theory. Their usefulness
stems mainly from their robustness---they are often easily computed even
in very strongly interacting theories and they are insensitive to a large variety of deformations
of the theory. Further, their phenomenological significance in fields ranging from solid state
physics to cosmology has inspired a large body of work in the last few decades devoted to
understanding the dynamics of anomalies from various viewpoints.

Despite this effort there is much that is yet to be understood. We have but a 
vague comprehension of the dynamics of anomalies at finite 
temperature and chemical potential, let alone far away from equilibrium. 
The existence of thermodynamic observables which are sensitive only to anomalies, or a classification of hydrodynamic processes which are responsive only to anomalies are examples  of problems which one would like to address.

While a complete understanding of the role of anomalies in terms of thermodynamic and hydrodynamic response is still elusive, a significant amount of progress has been made in the last few years using only a handful of tools and techniques. These tools have successfully explained various features of anomaly induced response in the case of pure $U(1)$ anomalies in hydrodynamics~\cite{Son:2009tf}. In what follows, we would like to argue that the existing methods which have been so successful in dealing with $U(1)$ anomalies are bound to give incomplete information when applied to gravitational or mixed anomalies. In order to do so, let us briefly summarize some of the salient features of hydrodynamics. 

The equations of motion of relativistic hydrodynamics are energy-momentum conservation and charge conservation when appropriate. These equations must be supplemented by constitutive relations which express the (covariant) energy momentum tensor $T^{\mu\nu}$ and current $J^{\mu}$ in terms of the hydrodynamic variables: $T$, $\mu$ and $u^{\mu}$---temperature, chemical potential and velocity field respectively. We normalize the velocity field such that $u^2=-1$. If we also turn on a slowly varying background gauge field $A_{\mu}$ and metric $g_{\mu\nu}$, then the constitutive relations will depend on gauge and diffeomorphism invariant combinations of the background fields. 

The constitutive relations can be decomposed into scalar, vector, and tensor relations with respect to the rotational symmetry which preserves the velocity field,
\begin{align}
\label{E:TJdecomp}
\begin{split}
J^{\mu} &= \mathcal{N} u^{\mu} + \nu^{\mu}\,, \\
T^{\mu\nu} &= \mathcal{E} u^{\mu} u^{\nu} + \mathcal{P} \Delta^{\mu\nu} + u^{\mu} q^{\nu} + u^{\nu}q^{\mu}+\tau^{\mu\nu}\,,
\end{split}
\end{align}
where we have defined the transverse projector $\Delta^{\mu\nu}=g^{\mu\nu}+u^{\mu}u^{\nu}$. The vectors $\nu^{\mu}$ and $q^{\mu}$ are transverse to $u^{\mu}$, and $\tau^{\mu\nu}$ is transverse and traceless. The constitutive relations~\eqref{E:TJdecomp} are usually arranged in a derivative expansion in which expressions which involve gradients of the hydrodynamic variables are suppressed by appropriate powers of the mean free path \cite{Baier:2007ix,Bhattacharyya:2008jc}. The constitutive relations characterize the response of the energy momentum tensor (or of the charge current) to a perturbation via response parameters. For instance, the conductivity parameterizes the response of the charge current to an external electric field. 

In some instances ``transport coefficients'' and ``response parameters'' are used interchangeably. In what follows we will make a distinction between transport coefficients or hydrodynamic response parameters, and thermodynamic response parameters (whereby response parameters or response coefficients will refer to either set). Transport coefficients characterize out of equilibrium dynamics and their Kubo formulae will involve correlators at non zero frequency. Thermodynamic response parameters are associated with the properties of the equilibrated system and the the appropriate Kubo formulae will involve correlators at zero frequency. The shear viscosity, bulk viscosity, and conductivity are examples of transport coefficients. An example of a thermodynamic response parameter is the magnetic susceptibility.

We are now in a position to discuss anomaly-induced transport as it is currently understood
\cite{Bhattacharyya:2007vs,Erdmenger:2008rm,Banerjee:2008th,Torabian:2009qk,Son:2009tf,Kharzeev:2009p,Lublinsky:2009wr,Neiman:2010zi,Bhattacharya:2011tra,Kharzeev:2011ds,Loganayagam:2011mu,Neiman:2011mj,Dubovsky:2011sk,Kimura:2011ef,Lin:2011aa,Nair:2011mk,Banerjee:2012iz,Jensen:2012jy,Jain:2012rh,Bhattacharyya:2012xi}.
In $1+1$ dimensions, we parameterize the anomalies through the conservation equations for the covariant current and stress tensor,
\begin{equation}
\label{E:2dnCov}
	\nabla_{\mu}J_{cov}^{\mu} = c_s \epsilon^{\mu\nu}F_{\mu\nu}\,,
	\qquad
	\nabla_{\nu}T_{cov}^{\mu\nu} = F^{\mu}{}_{\nu}J_{cov}^{\nu} + c_g \epsilon^{\mu\nu}\nabla_{\nu} R\,,
\end{equation}
where $F_{\mu\nu}$ is the field strength of the background gauge field, $R$ is the Ricci scalar of the background metric, and $\nabla$ indicates the covariant derivative using the Levi-Civita connection. To zeroth order in derivatives, the constitutive relations are given by
\begin{subequations}
\label{E:2dall}
\begin{align}
	\mathcal{P} &= P\,, 
	& \mathcal{E} &= -P + \mu \frac{\partial P}{\partial\mu}+T\frac{\partial P}{\partial T}\,, 
	& \mathcal{N} &= \frac{\partial P}{\partial \mu}\,,  \\
\label{E:2dcomponents}
	\nu^{\mu} & = -2 c_s\mu \epsilon^{\mu\nu}u_{\nu}\,,
	& q^{\mu} &= \epsilon^{\mu\nu}u_{\nu}(\tilde{c}_{2d}T^2 - c_s \mu^2)\,,
	& \tau^{\mu\nu} &=0\,,
\end{align}
\end{subequations}
where $P$ is the pressure and $\tau^{\mu\nu}$ identically vanishes in two dimensions \cite{Dubovsky:2011sk,Jain:2012rh}. The parameter $\tilde{c}_{2d}$ is a free constant which violates parity and time-reversal while preserving charge conjugation. 
At the one and two derivative level equation \eqref{E:2dall} receives several corrections which we will discuss in section \ref{S:generating2d}. The author of \cite{Valle:2012em} showed that in the presence of a  two dimensional gravitational anomaly the vector contribution $q^{\mu}$ to the constitutive relations \eqref{E:2dcomponents} becomes:\footnote{The gravitational anomaly 
coefficient $c_g$ is denoted by the symbol $-D$ in \cite{Valle:2012em}. Further we have used 
the following equilibrium relation to simplify the result derived in \cite{Valle:2012em} 
\begin{equation*} \prn{u^\alpha\nabla^\beta-u^\beta \nabla^\alpha}\nabla_\alpha u_\beta = -\frac{\nabla^2 T}{T}\,. \end{equation*}}
\begin{equation}
\label{E:cgcontribution}q^{\mu} = \epsilon^{\mu\nu} u_{\nu} \left( \tilde{c}_{2d} T^2 - c_s \mu^2 +2c_g\frac{\nabla_{\rho}\nabla^{\rho}T}{T}\right)\, .\end{equation}

In $3+1$ dimensions we parameterize a $U(1)^3$ and mixed gauge-gravitational anomaly by
\begin{align}
\begin{split}
	\nabla_{\mu}J^{\mu}_{cov}
	& = \frac{1}{4}\epsilon^{\mu\nu\rho\sigma}\left[ 3c_A F_{\mu\nu}F_{\rho\sigma} + c_m R^{\alpha}_{\phantom{\alpha}\beta\mu\nu}R^{\beta}_{\phantom{\beta}\alpha\rho\sigma}\right]\, \\
	\nabla_{\nu}T^{\mu\nu}_{cov}
	& = F^{\mu}_{\phantom{\mu}\nu}J_{cov}^{\nu} + \frac{c_m}{2}\nabla_{\nu}\left[\epsilon^{\rho\sigma\alpha\beta}F_{\rho\sigma}R^{\mu\nu}_{\phantom{\mu\nu}\alpha\beta}\right]\,.
\end{split}
\end{align}
To first order in derivatives, the constitutive relations are given by~\cite{Son:2009tf,Neiman:2010zi,Bhattacharya:2011tra}
\begin{subequations}
\label{E:4dconstRel}
\begin{equation}
	\mathcal{P}  = P - \zeta \nabla_{\mu}u^{\mu}\, ,
	\qquad
	\mathcal{E} = -P+\mu\frac{\partial P}{\partial \mu}+T\frac{\partial P}{\partial T}\, ,
	\qquad
	\mathcal{N} = \frac{\partial P}{\partial \mu}\, ,
\end{equation}
and 
\begin{align}
\begin{split}
	\nu^{\mu} &= -6c_A\mu B^{\mu} + (\tilde{c}_{4d}T^2-3c_A\mu^2)\omega^{\mu} + \sigma \Delta^{\mu\nu}\left(E_{\nu}-T\nabla_{\nu}\frac{\mu}{T} \right) \, , \\
	q^{\mu} & = (\tilde{c}_{4d}T^2-3c_A \mu^2)B^{\mu} + 2(\tilde{c}_{4d}\mu T^2 - c_A \mu^3)\omega^{\mu} \,, \\
	\tau^{\mu\nu} & = -\eta \sigma^{\mu\nu}\,,
\end{split}
\end{align}
where we have defined
\begin{align}
 	B^{\mu} & =\frac{1}{2}\epsilon^{\mu\nu\rho\sigma}u_{\nu}F_{\rho\sigma}\,,
	& \omega^{\mu} & = \epsilon^{\mu\nu\rho\sigma}u_{\nu}\nabla_{\rho}u_{\sigma}\, , \\
	E_{\mu} & =F_{\mu\nu}u^{\nu}\,, 
	& \sigma^{\mu\nu} & = \Delta^{\mu\rho}\Delta^{\nu\sigma}\left(\nabla_{\rho}u_{\sigma} + \nabla_{\sigma}u_{\rho}\right) - \frac{1}{3}\Delta^{\mu\nu}\nabla_{\rho}u^{\rho}\,
\end{align}
\end{subequations}
From the hydrodynamic analysis of~\cite{Neiman:2010zi}, the parameter $\tilde{c}_{4d}$ is a constant which violates parity and charge conjugation while respecting time-reversal. There is no  four dimensional analogue
of the two dimensional computation of \cite{Valle:2012em} in the literature. We carry out such an
analysis in this paper and our results are presented in appendix \ref{A:4dconstitutive}.

Explicit calculations in various $1+1$ dimensional CFTs with pure gravitational anomalies 
support the relation 
\begin{equation}
\label{E:2dCSA}
\tilde{c}_{2d} = - 8 \pi^2 c_g\,.
\end{equation}
The evidence for this statement in 
various $1+1$ dimensional CFTs is extensive, though the results  are somewhat fragmented across papers in 
various subfields of physics, ranging from thermal hall physics and superfluids
(See for example \cite{Volovik:1999wx,2002NuPhB.636..568C,2012PhRvB..85d5104R,2012PhRvB..85r4503S}
and references therein) 
through free field computations \cite{Landsteiner:2011cp,Loganayagam:2012pz}
to various studies of AdS$_3$/CFT$_2$ and thermodynamics of BTZ blackholes in the presence
of gravitational Chern-Simons terms \cite{Kraus:2005zm} (See \cite{Kraus:2006wn} for a review).
Similarly, computations in an arbitrary free field  theory of chiral fermions
\cite{Vilenkin:1978hb,Vilenkin:1979ui,Vilenkin:1980fu,Vilenkin:1980zv,Vilenkin:1980ft,Vilenkin:1995um,Loganayagam:2012pz,2012PhRvB..85d5104R} (see also \cite{Volovik:1999wx})
and in AdS$_5$/CFT$_4$ \cite{Landsteiner:2011iq} 
are consistent with the  relation
\begin{equation}
\label{E:4dCSA}
\tilde{c}_{4d} = - 8 \pi^2 c_m\,,
\end{equation}
in four dimensions.
A natural generalisation of~\eqref{E:2dCSA} and~\eqref{E:4dCSA} to arbitrary dimensions
was presented in~\cite{Loganayagam:2012pz} by an analysis of free fermion theories.

The peculiar relations \eqref{E:2dCSA} and \eqref{E:4dCSA} are the main focus of this paper. 
They exhibit two distinct features which {a priori} indicate that a general proof of their validity (if it exists) will require special attention.
\begin{enumerate}
\item{\emph{A breakdown of the naive derivative expansion}: The gravitational and mixed anomalies contribute to the response parameters at two orders of derivatives lower than expected when studying the equations of motion. We will refer to this phenomenon as a ``jump'' in the derivative expansion. For instance, the gravitational anomaly in $1+1$ dimensions enters at zeroth order in the derivative expansion instead of second order in the derivative expansion. 
Existing methods~\cite{Son:2009tf,Banerjee:2012iz,Jensen:2012jh} independently constrain the constitutive relations order by order in derivatives, and so cannot exhibit jumps in the derivative expansion. Indeed, when applied to the case of a two dimensional theory with a gravitational anomaly~\cite{Valle:2012em}, existing methods lead to the contribution~\eqref{E:cgcontribution} to the constitutive relations, but fail to establish~\eqref{E:2dCSA}.
}
\item{\emph{The $8\pi^2$ factor}: All existing methods to derive constraints on anomaly induced transport~\cite{Son:2009tf,Neiman:2010zi,Bhattacharya:2011tra,Banerjee:2012iz,Jensen:2012jh} carry out algebraic manipulations on the anomalous conservation equations (for example \eqref{E:2dnCov}). 
Using our normalization for the various coefficients, a moment of thought reveals that the constraints derived out of such algebraic arguments cannot contain transcendental numbers like the additional factor of  $8\pi^2$ in \eqref{E:2dCSA}. This obstruction by itself precludes any constraint of the form~\eqref{E:2dCSA} to arise from methods based on \cite{Son:2009tf}. Moreover, any argument which gives a constraint such as \eqref{E:2dCSA} or \eqref{E:4dCSA} is expected to be {geometric} rather than algebraic. 
}
\end{enumerate}

Fortunately, there is a very well-known relation in $1+1$ dimensional conformal field theory (2d CFT) which exhibits the same features as \eqref{E:2dCSA} or \eqref{E:4dCSA}---the Cardy formula. Given a unitary modular-invariant 2d CFT with a Weyl anomaly
\begin{equation}
\label{eq:cwCFT}
(T_{cov})^\mu_\mu =  c_w R = \frac{1}{2\pi}\frac{c_R+c_L}{24} R \,,
\end{equation}
the Cardy formula relates the pressure of this CFT in the infinite volume limit to the left and right handed central charges $c_R$ and $c_L$ via~\cite{Bloete:1986qm,Affleck:1986bv}
\begin{equation}
\label{E:Peqc} 
P =  2\pi T^2\frac{c_R+c_L}{24} = 4\pi^2 c_w T^2\,.
\end{equation}
The relation \eqref{E:Peqc} follows, essentially, from the Casimir energy picked up when going from two-dimensional Euclidean space to the cylinder due to the non-conformal behavior of the energy momentum tensor. The relation~\eqref{E:Peqc} exemplifies a relation between terms of different orders in the derivative expansion. (In this case, the pressure is zeroth order in the derivative expansion while the Weyl anomaly is second order.) Further, the transcendental number here $4\pi^2$ comes essentially out of the geometry of the thermal circle; $2\pi T$ is precisely the inverse radius of the cylinder.
In fact, as we show in appendix \ref{A:Casimir}, one can easily extend the result \eqref{E:Peqc} to theories with $c_g \neq 0$ therey deriving \eqref{E:2dCSA} for conformal field theories with no charge current.

In the rest of this work, we use the intuition gained by the observation~\eqref{E:Peqc} to demonstrate~\eqref{E:2dCSA} and~\eqref{E:4dCSA} by generalizing some recently obtained results for the Euclidean generating functional~\cite{Banerjee:2012iz,Jensen:2012jh}. We will argue that the jumps in the derivative expansion in $1+1$ and $3+1$ dimensional theories can be understood in terms of a ``Casimir momentum density'' which can be computed by placing the theory on a cone (or a product of a cone with a two dimensional manifold). We also compute the constitutive relations of a conformal and non-conformal theory in $1+1$ dimensions to second order in the derivative expansion (see~\eqref{E:general}), as well as the contribution of the mixed anomaly to the constitutive relations in $3+1$ dimensions (which may be found in appendix~\ref{A:4dconstitutive}).

The remainder of this manuscript is organized as follows. In sections~\ref{S:anomalies} and \ref{S:generating} we collect various results from the literature relevant to this work: section~\ref{S:anomalies} briefly reviews the structure of the anomaly polynomial and the construction of the Bardeen-Zumino anomaly polynomial which distinguishes between covariant and consistent observables. 
Section~\ref{S:generating} summarizes the method of construction of an equilibrium partition function~\cite{Banerjee:2012iz,Jensen:2012jh} which will play an important role in our work. The reader familiar with the content of sections \ref{S:anomalies}--\ref{S:generating} may go directly to sections~\ref{S:2d} and~\ref{S:4d} which contain our main result. In section~\ref{S:2d} we compute the effect of a gravitational anomaly on the response coefficients of two dimensional theories and in section~\ref{S:4d} we compute the effect of a mixed anomaly on the response coefficients of four dimensional theories. We end with a discussion and summary in section~\ref{S:SandD}.

\emph{Note:} While this work was nearing completion, we were made aware of~\cite{UW:CVE} which deals with issues similar to the ones considered in this paper.

\section{Anomalies}
\label{S:anomalies}
In this section we briefly review several useful facts about gauge and
gravitational  anomalies and establish our notation. Since these 
facts are well known, we will necessarily be brief. 
We will refer the reader to various books and lecture notes
\cite{BBook,Harvey:2005it,Bilal:2008qx} for a more detailed review. 
The reader familiar with these results may skip to Section~\ref{S:generating}. 

\subsection{ Anomaly polynomial, Consistent and covariant anomalies}
\label{S:AP}
Consider a $2n$-dimensional theory with a conserved $U(1)$ current and stress
tensor which respectively couple to a background gauge field $A$ and metric $g$.
We will assume that these conservation laws become anomalous once we place the
theory on a non-trivial background. 

We will denote the curvature of $A$ by $F=dA$, and  we define the Christoffel
connection 1-forms and curvature 2-forms associated  with the metric via
\begin{equation}
\Gamma^{\alpha}{}_{\beta}\equiv \Gamma^{\alpha}{}_{\beta\mu}dx^{\mu}, \qquad 
\mathfrak{R}^{\alpha}{}_{\beta}\equiv
\frac{1}{2}R^{\alpha}{}_{\beta\mu\nu}dx^{\mu}\wedge dx^{\nu}
=d\Gamma^{\alpha}{}_{\beta}+\Gamma^{\alpha}{}_{\gamma}\wedge
\Gamma^{\gamma}{}_{\beta},
\end{equation}
where Greek indices are  spacetime indices, $\Gamma^{\alpha}{}_{\beta\mu}$ are
the Levi-Civita connection coefficients, and $R^{\alpha}{}_{\beta\mu\nu}$ are 
the components of the Riemann tensor. 

Let $W$ denote the generating function of the $2n$ dimensional theory 
we are interested in. Let us also define a $2n+1$ dimensional theory on 
a manifold  $\mathcal{M}_{2n+1}$ with  boundary $\partial \mathcal{M}_{2n+1}$
with a generating function $W_{cov}$ defined through
\begin{equation}
\label{E:defWcov}
	W_{cov} =  W[\partial\mathcal{M}_{2n+1}] +  \int_{\mathcal{M}_{2n+1}}I^{CS}_{2n+1}\,.
\end{equation}
Here $I^{CS}_{2n+1}$ is a Chern-Simons form associated with the metric, 
gauge field, or a combination of the two. The form of the 
$2n+1$ dimensional  Chern-Simons term is determined by demanding
that the covariant generating function, $W_{cov}$, is invariant 
under gauge transformations and diffeomorphisms. More precisely if we 
denote gauge and diffeomorphism variation by $\delta_{\lambda}$, 
\begin{equation}\label{E:deltaAg}
\begin{split}
\delta_{\lambda} A_{\mu} &= \partial_{\mu} \Lambda + A_{\nu}\partial_{\mu}
\xi^{\nu} + (\partial_{\nu} A_{\mu})\xi^{\nu} = \partial_{\mu} \prn{\Lambda +A_{\nu} \xi^{\nu}}+\xi^\nu F_{\nu\mu},\\
\delta_{\lambda} g_{\mu\nu} &= \nabla_{\mu} \xi_{\nu} + \nabla_{\nu} \xi_{\mu}\,,
\end{split}
\end{equation}
then the statement that the $2n+1$ dimensional theory is gauge and diffeomorphism invariant implies that $\delta_{\lambda}W_{cov}=0$.

As a result of the gauge invariance of $W_{cov}$,
the $2n$ dimensional theory described by the generating function $W$ in \eqref{E:defWcov} is 
not gauge and diffeomorphism invariant and hence, anomalous.
In detail, the  gauge and diffeomorphism invariance of $W_{cov}$ amounts to 
\begin{equation}\label{E:deltaWcov}
	\delta_{\lambda} W[\partial\mathcal{M}_{2n+1}] +  \delta_{\lambda} \int_{\mathcal{M}_{2n+1}}I^{CS}_{2n+1} = 0\,.
\end{equation}
By construction, the bulk Chern-Simons term is gauge-invariant up to boundary contributions. Let us
parameterize the variation of the Chern-Simons term by
\begin{equation}\label{E:deltaCS}
\begin{split}
	\delta_\lambda \int_{\mathcal{M}_{2n+1}}I^{CS}_{2n+1} 
	&= \int_{\partial\mathcal{M}_{2n+1}} d^{2n}x\sqrt{-g}\  \ \Lambda\ \mathcal{J}
	+ \int_{\partial\mathcal{M}_{2n+1}} d^{2n}x\sqrt{-g}\ \partial_{\lambda}\xi^{\nu}\ \mathcal{T}^\lambda{}_\nu \,,
\end{split}
\end{equation}
so that \eqref{E:deltaWcov} reads 
\begin{equation}\label{E:deltaWanom}
	\delta_{\lambda} W[\mathcal{M}_{2n}] = - \int_{\mathcal{M}_{2n}} d^{2n}x\sqrt{-g}\  \ \Lambda\ \mathcal{J}
	- \int_{\mathcal{M}_{2n}} d^{2n}x\sqrt{-g}\ \partial_{\lambda}\xi^{\nu}\ \mathcal{T}^\lambda{}_\nu  \neq 0.
\end{equation}

The fact that the generating function $W$ is not gauge and (or) diffeomorphism invariant signals
the presence of anomalies. From the viewpoint of the $2n+1$ dimensional theory, 
the $2n$ dimensional anomaly is due to an anomaly inflow---a flow of conserved charges 
from the bulk to the boundary thereby modifying the conservation laws at the boundary.

Let us consider these conservation laws in some detail. 
We define a consistent stress tensor and a consistent charge current 
by varying the generating function $W$ with respect to the metric and gauge field
respectively.\footnote{ The ``consistency'' of the current and stress tensor 
are associated with the so called Wess-Zumino consistency conditions \cite{Wess:1971yu} (See \cite{BBook,Harvey:2005it,Bilal:2008qx}
for a review).}
\begin{equation}\label{E:TJCons}
\begin{split}
	J^{\mu}_{cons} &\equiv \frac{1}{\sqrt{-g}}\frac{\delta W}{\delta A_{\mu}}, \\
	T^{\mu\nu}_{cons}&\equiv \frac{2}{\sqrt{-g}}\frac{\delta W}{\delta g_{\mu\nu}}\,.
\end{split}
\end{equation}
We can now compute the gauge  and  diffeomorphism variation of $W$, 
\begin{equation}\label{E:deltaWTJ}
\begin{split}
\delta_{\lambda} W &= \int d^{2n}x \left[ \frac{\delta W}{\delta
g_{\mu\nu}}\delta_{\lambda}g_{\mu\nu}+\frac{\delta W}{\delta
A_{\mu}}\delta_{\lambda}A_{\mu}\right], \\
 &= \int d^{2n}x \sqrt{-g}\brk{- \Lambda
(\nabla_{\mu} J_{cons}^{\mu}) -\xi_{\mu}\left(\nabla_{\nu} T^{\mu\nu}_{cons}
-F^{\mu}{}_\nu J_{cons}^\nu + A^{\mu} \nabla_{\nu} J^{\nu}_{cons}\right)  }, 
\end{split}
\end{equation}
where in going from the first line to the second we have integrated by parts. Here
$\nabla_{}$ represents the covariant differentiation of tensors using the 
usual Christoffel connection. Generically, in a theory with diffeomorphism anomalies none
of the currents are tensors---we define 
the action of $\nabla_{}$ on $T_{cons}^{\mu\nu}$ and $J_{cons}^{\mu}$ by treating 
them as if they were tensorial objects. 
Comparing \eqref{E:deltaWTJ} against \eqref{E:deltaWanom} we get the {consistent} anomaly equations:
\begin{equation}\label{eq:AnomCons}
\begin{split}
	\nabla_{\mu} J^{\mu}_{cons} &= \mathcal{J}, \\
	\nabla_{\nu} T^{\mu\nu}_{cons} &= F^{\mu}_{\phantom{\mu}\nu}J_{cons}^{\nu} 
	-\mathcal{J} A^\mu
	- g^{\mu\nu}\frac{1}{\sqrt{-g}}\ \partial_\lambda \brk{\sqrt{-g}\ \mathcal{T}^\lambda{}_\nu } \,.
\end{split}
\end{equation}
We emphasize that neither the currents $J^{\mu}_{cons}$ and $T^{\mu\nu}_{cons}$ nor 
the anomalies on the right hand side of equation \eqref{eq:AnomCons} 
transform covariantly under gauge transformations and diffeomorphisms. 
This unfortunate state of affairs may be remedied by using 
the generating function $W_{cov}$ instead of $W$. 

Consider a general variation of the $2n+1$ dimensional metric $g^{(2n+1)}_{ab}$ 
and the $2n+1$ dimensional gauge field $A^{(2n+1)}_{a}$ on $\mathcal{M}_{2n+1}$. 
In what follows Roman indices will be associated with the coordinates on 
$ \mathcal{M}_{2n+1}$ and Greek indices with the coordinates on 
$\partial \mathcal{M}_{2n+1}$. The index associated with the 
direction normal to the boundary will be denoted by $\bot$.
The orientation of the normal direction is fixed by demanding
\begin{equation}
\sqrt{-g^{(2n+1)}} \epsilon_{2n+1}^{\bot \mu_1\ldots \mu_{2n} } =
\sqrt{-g} \epsilon^{ \mu_1\ldots \mu_{2n} }\,.
\end{equation}
The variation of the $2n+1$ dimensional metric will contain 
a bulk piece $\delta g_{ab}$ and a boundary piece $\delta g_{\mu\nu}$.
Likewise, the variation of the gauge field will contain a bulk piece 
$\delta A_a$ and a boundary piece $\delta A_{\mu}$. With this notation
in mind, we can parameterize the variation of $W_{cov}$ with respect
to the sources through
\begin{multline}
\label{E:deltaSCSdef} 
	\delta_S \int_{\mathcal{M}_{2n+1}}I^{CS}_{2n+1}
	= \int_{\mathcal{M}_{2n+1}}d^{2n+1}x\sqrt{-g^{(2n+1)} }  \Bigl\{   \mathrm{J}^a\delta A_a 
	+ \mathrm{L}^{ab}{}_c \delta\Gamma^c{}_{ab} \Bigr\} \\
	+ \int_{\partial\mathcal{M}_{2n+1}}  d^{2n}x\sqrt{-g} \Bigl\{Y^\mu \delta A_\mu 
	+ \frac{1}{2} X^{\lambda\mu}{}_\nu \delta\Gamma^\nu{}_{\lambda\mu}\Bigr\} \,.
\end{multline}

Before we proceed, certain clarifying remarks are in order. The metric variation  
enters entirely through the variation of the connection because of the 
topological nature of the Chern-Simons terms. Furthermore, we have 
ignored all dependencies on, say, the extrinsic curvature of the boundary 
since it is not relevant to the discussion below.\footnote{
	This involves the following assumption. The contractions in the boundary terms
involve only the sum over the boundary indices. While this is sufficient for
our purposes, we warn the reader that a more careful treatment of such extrinsic 
terms is required in many other instances. For example when carrying out computations in the framework of the AdS/CFT 
correspondence such terms (along with other Gibbons-Hawking like terms)
give a finite gauge(or diffeomorphism)-invariant
contribution to the dual stress tensor~\cite{Kraus:2005zm}.
}
Since the Christoffel symbols are symmetric in their last two indices, 
the tensors $\mathrm{L}$ and $X$ are well defined only up to 
tensors which are anti-symmetric in their first two indices. In
what follows we will fix this ambiguity by demanding that
\begin{equation}
\begin{split}
	\mathrm{L}_{abc} & = -\mathrm{L}_{acb}, \\
	X^{\mu\lambda}{}_\nu &=X^{\lambda\mu}{}_\nu.\\
\end{split}
\end{equation}

Converting the variation of the Christoffel symbols in \eqref{E:deltaSCSdef} to metric variations
we can write
\begin{multline}
\label{E:deltaSCS}
	\delta_S \int_{\mathcal{M}_{2n+1}}I^{CS}_{2n+1}
	= \int_{\mathcal{M}_{2n+1}}d^{2n+1}x\sqrt{-g_{_{2n+1}} }  \Bigl\{  \mathrm{J}^a\delta A_a 
	+\nabla_c \mathrm{L}^{abc} \delta g_{ab} \Bigr\} \\
	\qquad + \int_{\partial\mathcal{M}_{2n+1}}  d^{2n}x\sqrt{-g} \Bigl\{P^\mu_{BZ} \delta A_\mu 
	+ \frac{1}{2} P_{BZ}^{\mu\nu}\delta g_{\mu\nu}\Bigr\},
\end{multline}
with
\begin{equation}\label{eq:defPX}
\begin{split}
	P^{\mu}_{BZ} = Y^{\mu},
	\qquad
	P^{\mu\nu}_{BZ} \equiv
	-\frac{1}{2} \nabla_{\lambda}\prn{ 
	X^{\lambda\mu\nu}+X^{\lambda\nu\mu}-X^{\mu\nu\lambda}},
\end{split}
\end{equation}
and Greek indices are raised and lowered with the metric $g_{\mu\nu}$. In the equations 
above we have again dropped extrinsic contributions to $P^{\mu\nu}_{BZ}$ (of
the form $\mathrm{L}^{\mu\bot\nu}+\mathrm{L}^{\nu\bot\mu}$ coming from an integration
by parts) .

We are now in a position to define the covariant currents as
the variation of $W_{cov}$ with respect to the gauge fields and metric on $\partial \mathcal{M}_{2n+1}$,
\begin{equation}\label{E:covJT}
\begin{split}
J^{\mu}_{cov} &\equiv \frac{1}{\sqrt{-g}}\frac{\delta W_{cov}}{\delta A_{\mu}} = J^{\mu}_{cons} +  P^\mu_{BZ}, \\
T^{\mu\nu}_{cov}&\equiv \frac{2}{\sqrt{-g}}\frac{\delta W_{cov}}{\delta g_{\mu\nu}}=  T^{\mu\nu}_{cons} +  P^{\mu\nu}_{BZ}\,.
\end{split}
\end{equation}
By construction these currents transform covariantly under gauge transformations and diffeormorphisms 
(they were obtained by varying an invariant functional). In other words, the non-covariant 
transformations of $P_{BZ}^{\mu}$ and $P_{BZ}^{\mu\nu}$ under gauge transformations and diffeomorphisms
exactly compensates for the non covariant transformation properties of the consistent currents, implying
that the covariant currents transform covariantly as advertised. The corrections $P^\mu_{BZ}$ and $P^{\mu\nu}_{BZ}$ 
that  come from varying the bulk Chern-Simons terms are are often termed the {Bardeen-Zumino currents} 
or Bardeen-Zumino polynomials \cite{Bardeen:1984pm} which is the reason for the subscript `$BZ$' in our notation.

Let us turn to the conservation laws for the covariant currents. To compute these 
conservation laws we need to express the divergence of the Bardeen-Zumino currents in terms of $\mathcal{T}^{\mu}{}_\nu$ 
and $\mathcal{J}$ defined in \eqref{E:deltaCS}. This computation can be carried out
by using \eqref{E:deltaSCS} to evaluate the gauge and diffeomorphism variation of $\int I^{CS}_{2n+1}$. 
Since the variation of the Chern-Simons term gets contributions only from the boundary we find,
\begin{multline}
\label{E:deltaCSBZ}
	\delta_\lambda \int_{\mathcal{M}_{2n+1}}I^{CS}_{2n+1}\\
	=  -\int_{\partial\mathcal{M}_{2n+1}} d^{2n}x \sqrt{-g}\ \xi_{\mu}\left(\nabla_{\nu} P_{BZ}^{\mu\nu}-\nabla_\nu  \mathrm{L}^{\bot\mu\nu} -F^\mu{}_\nu P^\nu_{BZ} + A^{\mu} \brk{\nabla_{\nu} P_{BZ}^{\nu}-\mathrm{J}^{\bot}} \right) \\
	-\int_{\partial\mathcal{M}_{2n+1}} d^{2n}x \sqrt{-g}\ \Lambda (\nabla_{\mu} P_{BZ}^{\mu}-\mathrm{J}^{\bot})\,.
\end{multline}
Comparing \eqref{E:deltaCSBZ} against \eqref{E:deltaCS} we get
\begin{equation}\label{E:divBZ}
\begin{split}
\nabla_{\mu} P_{BZ}^{\mu} &= \mathrm{J}^{\bot}-\mathcal{J},\\
\nabla_{\nu} P_{BZ}^{\mu\nu} &=F^\mu{}_\nu P^\nu_{BZ} 
+ \mathcal{J} A^{\mu} +\nabla_\nu  \mathrm{L}^{\bot\mu\nu}
+ g^{\mu\nu}\frac{1}{\sqrt{-g}}\ \partial_\lambda \brk{\sqrt{-g}\ \mathcal{T}^\lambda{}_\nu }\,.
\end{split}
\end{equation}
Finally, combining \eqref{E:divBZ} with \eqref{eq:AnomCons} we find
\begin{equation}\label{eq:Anomcov}
\begin{split}
\nabla_{\mu} J^{\mu}_{cov} &= \mathrm{J}^{\bot}, \\
\nabla_{\nu} T^{\mu\nu}_{cov} &= F^\mu{}_\nu J^\nu_{cov} 
+\nabla_\nu  \mathrm{L}^{\bot\mu\nu} \,.
\end{split}
\end{equation}

An alternate way of deriving the conservation law \eqref{eq:Anomcov} is to begin with 
the variation of $W_{cov}$ with respect to an arbitrary variation in sources
\begin{multline}
\label{E:deltaSWcov}
\delta_S W_{cov}
= \delta_S W[\partial\mathcal{M}_{2n+1}] + \delta_S \int_{\mathcal{M}_{2n+1}}I^{CS}_{2n+1}\\
= \int_{\mathcal{M}_{2n+1}}d^{2n+1}x\sqrt{-g_{_{2n+1}} }  \Bigl\{  \mathrm{J}^a\delta A_a 
+\nabla_c \mathrm{L}^{abc} \delta g_{ab} \Bigr\}  \\
+ \int_{\partial\mathcal{M}_{2n+1}}  d^{2n}x\sqrt{-g} \Bigl\{J^\mu_{cov} \delta A_\mu 
+ \frac{1}{2} T_{cov}^{\mu\nu}\delta g_{\mu\nu}\Bigr\},
\end{multline}
and then demand that the gauge and diffeomorphism variation of $W_{cov}$ vanish.
From this derivation it is clear that the covariant anomaly equations 
\eqref{eq:Anomcov} can be thought of as the conservation laws for
the covariant charge applied to $\mathcal{M}_{2n+1}$ and its boundary. 
This anomaly inflow phenomenon is often called the Callan-Harvey anomaly inflow 
mechanism\cite{Callan:1984sa,Naculich:1987ci,Harvey:2000yg,Harvey:2005it}.

While not relevant for this work, before we move on to particular examples, we note that we can always define
a canonically conserved energy-momentum tensor 
\[ {T}_{conserved}^{\mu\nu} \equiv T^{\mu\nu}_{cov}-\mathrm{L}^{\bot\mu\nu}, \]
such that
\begin{equation}\label{eq:AnomcovLorentz}
\begin{split}
\nabla_{\mu} J^{\mu}_{cov} &= \mathrm{J}^{\bot}, \\
\nabla_{\nu} {T}^{\mu\nu}_{conserved} &= F^{\mu}{}_{\nu}J_{cov}^{\nu}, \\
{T}^{\mu\nu}_{conserved}- {T}^{\nu\mu}_{conserved} &=- 2 \mathrm{L}^{\bot\mu\nu}, \\
\end{split}
\end{equation}
where the last equation can be interpreted as the Lorentz anomaly---the non-conservation
of the angular momentum current. Thus, $-2\mathrm{L}^{\bot\mu\nu}$ 
can be thought of as the rate at which  angular momentum is injected into the system. 
This demonstrates the well-known result in the theory of gravitational anomalies: we can 
always trade a diffeomorphism anomaly for a Lorentz anomaly.

Note that to study gravitational anomalies one needs to necessarily place the $2n$ dimensional theory 
on a non-Minkowski background without Poincare invariance and hence we need to clarify 
what we mean by `angular momentum' in the discussion above. Consider a Killing vector which generates a symmetry of the given gauge and gravitational background. More precisely, we define a vector $K^\mu$ and a corresponding $U(1)$ gauge 
transformation parameter $\Lambda_K$ such that
\begin{equation}\label{eq:Killing}
\begin{split}
\pounds_K g_{\mu\nu} &= \nabla_\mu K_\nu + \nabla_\nu K_\mu =0,\\
\pounds_K A_\mu + \nabla_\mu \Lambda_K &= K^\nu F_{\nu\mu} + \nabla_\mu \prn{ K^\nu A_\nu + \Lambda_K}=0\,.
\end{split}
\end{equation}
According to Noethers theorem there exists a Noether Current
\begin{equation}\label{eq:Killing2}
\begin{split}
J^\mu_K &= K_\nu T^{\nu\mu}_{conserved}+ (K^\nu A_\nu + \Lambda_K)\ J^\mu_{cov},
\end{split}
\end{equation}
whose diveregence can be worked out using \eqref{eq:AnomcovLorentz} to be 
\begin{equation}\label{eq:Noether}
\begin{split}
\nabla_\mu J^\mu_K &= \frac{1}{2} T^{\nu\mu}_{conserved}\pounds_K g_{\mu\nu}+\prn{\pounds_K A_\mu + \nabla_\mu \Lambda_K} J^\mu_{cov}\\
&\qquad -\frac{1}{2}\mathrm{L}^{\bot\mu\nu}(\nabla_\mu K_\nu-\nabla_\nu K_\mu) +  \mathrm{J}^{\bot}\ (K^\nu A_\nu + \Lambda_K).\
\end{split}
\end{equation}
The first line of the above equation vanishes because of our assumption in \eqref{eq:Killing}
whereas for the second line to vanish either the background should have the anomalies turned off i.e.,  
$\mathrm{J}^{\bot}=0$ and $\mathrm{L}^{\bot\mu\nu}=0$
or an extra restriction needs to be satisfied by $\Lambda_K$ and $K^\mu$, i.e.,
$\nabla_\mu K_\nu-\nabla_\nu K_\mu=0$ and $K^\nu A_\nu + \Lambda_K =0$. 

Thus, the Noether currents corresponding to curl-free Killing vectors 
(`momentum-like Noether charges') can still  be conserved 
in the presence of gravitational anomalies whereas the  
Noether currents corresponding to Killing vectors which 
are \emph{not} curl-free  (`angular momentum-like Noether charges')  
get a contribution to their divergence proportional to 
$\mathrm{L}^{\bot\mu\nu}$. This is the precise meaning of 
the statement that the theory placed on a 
non-trivial gauge and gravitational background exhibits a Lorentz anomaly 
given by $-2\mathrm{L}^{\bot\mu\nu}$.

So far, we have characterized gauge and diffeomorphism anomalies in $2n$ dimensions using
a $2n+1$ dimensional Chern Simons form. The most convenient way to characterize $2n+1$ dimensional Chern-Simons forms is to construct a $2n+2$ form 
\begin{equation}
\mathcal{P}_{anom}=dI^{CS}_{2n+1}.
\end{equation}
which we call the anomaly polynomial of the theory. This anomaly 
polynomial is an index-like object built out of the Chern classes of $F$ 
and Pontryagin classes of $\mathfrak{R}$. Given an anomaly polynomial $\mathcal{P}_{anom}$
encoding the anomalies of a theory, one can reconstruct the Chern-Simons form
$I^{CS}_{2n+1}$ and from it the anomalous behavior of the stress tensor and current.

\subsection{Two dimensional theories}
\label{S:2danomalies}

The anomaly polynomial of an arbitrary theory in $d=2$ is a 4-form given by
\begin{equation}\label{E:anomP2d}
\mathcal{P}_{anom}[F,\mathfrak{R}] = c_s F\wedge F - 8\pi^2 c_g p_{_1}(\mathfrak{R}) = c_s F\wedge F + c_g
\text{tr}(\mathfrak{R} \wedge \mathfrak{R}) ,
\end{equation}
where $p_1$ is the first Pontryagin class
\begin{equation}
p_{_1}(\mathfrak{R}) \equiv -\frac{1}{8\pi^2}\text{tr}(\mathfrak{R} \wedge \mathfrak{R}) \,,
\end{equation}
and $\text{tr}$ denotes a trace over the Lorentz indices. The coefficient of the pure $U(1)^2$ Schwinger anomaly is denoted $c_s$ and $c_g$ is the 
coefficient associated with the pure gravitational anomaly. 

Given an arbitrary theory with chiral bosons and (or) fermions, $c_s$ and $c_g$ in the anomaly polynomial can 
be computed using
\begin{equation}
\begin{split}
c_s &= - \frac{2\pi}{2!(2\pi)^2}\sum_{i=species} \chi_i \prn{q_i^2}_{1/2},\\ 
c_g &= - \frac{2\pi}{4!(8\pi^2)}\sum_{i=species} \chi_i  \prn{1_0+1_{1/2}},  \\
\end{split}
\end{equation}
where $\chi_i$ denotes the chirality (we assign right handed fermions positive chirality), 
$q$ denotes the charge and the subscripts $\{0,1/2\}$ represent the
contribution from a chiral scalar and a Weyl-fermion respectively. The contribution of a
Majorana-Weyl fermion is half that of a Weyl-fermion of the same chirality. The sum above is
performed over each species where each particle/anti-particle pair contributes one term to the sum. 
In particular for a 2d-CFT with a $U(1)$ symmetry, we have
\begin{equation}
\label{E:csandcg}
\begin{split}
c_s &= - \frac{2\pi}{(2\pi)^2}(k_R-k_L),\\ 
c_g &= - \frac{2\pi}{4!(8\pi^2)}(c_R-c_L), \\
\end{split}
\end{equation}
where $k_{R/L}$ are the right/left $U(1)$ Kac-Moody levels and $c_{R/L}$ are the right/left Virasoro
central charges.

The Chern-Simons form corresponding to the anomaly polynomial \eqref{E:anomP2d} is
\begin{equation}
I^{CS}_3 = c_s A \wedge F +c_g j_{CS}. 
\end{equation}
where we have defined the gravitational Chern-Simons 3-form
\begin{equation}\label{eq:gravCS}
\begin{split}
 j_{CS}&\equiv \text{tr}\brk{ \Gamma\wedge d\Gamma +\frac{2}{3}\Gamma\wedge\Gamma\wedge \Gamma },\\
\end{split}
\end{equation}
such that $dj_{CS}=\text{tr}(\mathfrak{R} \wedge \mathfrak{R})$.
The gauge and diffeomorphism transformation of these Chern-Simons terms is given by
\begin{equation}
\label{E:CS3variation}
\delta_\lambda \int_{\mathcal{M}_3}I^{CS}_3 
= \int_{\partial\mathcal{M}_3}   \brk{c_s \Lambda F  + c_g (\partial_{\lambda}\xi^{\nu}) d\Gamma^{\lambda}{}_\nu } \,.
\end{equation}
Comparing \eqref{E:CS3variation} with \eqref{E:deltaCS} we get
\begin{align}\label{E:JT2d}
\begin{split}
\mathcal{J} &= c_s\ \frac{1}{2}\epsilon^{\alpha\beta}F_{\alpha\beta}\,, \\
\mathcal{T}^\lambda{}_\nu &= c_g \epsilon^{\alpha\beta}\partial_\alpha \Gamma^\lambda{}_{\nu\beta} \,.
\end{split}
\end{align}
Thus,
\begin{equation}
\begin{split}
\nabla_{\mu} J^{\mu}_{cons} &= c_s\ \frac{1}{2}\epsilon^{\alpha\beta}F_{\alpha\beta}, \\
\nabla_{\nu} T^{\mu\nu}_{cons} &= F^{\mu\nu}J^{cons}_{\nu} 
-c_s\ \frac{1}{2}\epsilon^{\alpha\beta}F_{\alpha\beta} A^\mu
- c_g\ g^{\mu\nu}\frac{1}{\sqrt{-g}}\ \partial_\lambda \brk{\sqrt{-g}\epsilon^{\alpha\beta}\partial_\alpha \Gamma^\lambda{}_{\nu\beta} }\,.
\end{split}
\end{equation}

To obtain the covariant current we vary the Chern-Simons forms with respect to the sources,
\begin{equation}
\label{E:ICS3sources}
	\delta_S \int_{\mathcal{M}_3}I^{CS}_3= \int_{\mathcal{M}_3}  \Bigl\{ 2 c_s \delta A\wedge F  + 2 c_g \text{tr}(\delta\Gamma \wedge \mathfrak{R}) \Bigr\} 
	+ \int_{\partial\mathcal{M}_3}   \Bigl\{ c_s \delta A \wedge  A   + c_g \delta\Gamma^\nu{}_\lambda \wedge \Gamma^\lambda{}_\nu \Bigr\}\,.
\end{equation}
Comparing \eqref{E:ICS3sources} against \eqref{E:deltaSCSdef}  we can read off the
covariant anomaly and the Bardeen-Zumino currents 
\begin{equation}
\begin{split}
  \mathrm{J}^{\bot} &=   2c_s\ \frac{1}{2}\epsilon^{\alpha\beta}F_{\alpha\beta},\\
 \mathrm{L}^{\bot\mu\nu} &= 2 c_g \frac{1}{2}\epsilon^{\alpha\beta}R^{\mu\nu}{}_{\alpha\beta},  \\
 P_{BZ}^\mu &= c_s \epsilon^{\mu\nu}A_{\nu},\\
X^{\mu\lambda}{}_\nu &= c_g  \epsilon^{\mu\rho}\Gamma^\lambda{}_{\nu\rho} +c_g  \epsilon^{\lambda\rho}\Gamma^\mu{}_{\nu\rho},\\
P^{\mu\nu}_{BZ} &=
-\frac{1}{2} \nabla_{\lambda}\prn{ 
X^{\lambda\mu\nu}+X^{\lambda\nu\mu}-X^{\mu\nu\lambda}},\\
\end{split}
\end{equation}
The covariant anomaly equations \eqref{eq:Anomcov} become
\begin{equation}
\begin{split}
\nabla_{\mu} J^{\mu}_{cov} &= 2c_s\ \frac{1}{2}\epsilon^{\alpha\beta}F_{\alpha\beta}, \\
\nabla_{\nu} T^{\mu\nu}_{cov} &= F^{\mu\nu}J^{cov}_{\nu} 
+2 c_g\ \nabla_\nu \brk{\frac{1}{2}\epsilon^{\alpha\beta}R^{\mu\nu}{}_{\alpha\beta} }.\\
\end{split}
\end{equation}
In two dimensions the Riemann tensor satisfies an identity 
\[ \epsilon^{\alpha\beta}R^{\mu\nu}{}_{\alpha\beta} = \epsilon^{\mu\nu} R \]
which can be used to bring the above equations to a more familiar form
\begin{equation}\label{E:2dcovCons}
\begin{split}
\nabla_{\mu} J^{\mu}_{cov} &= 2c_s\ \epsilon^{\alpha\beta}\partial_\alpha A_\beta, \\
\nabla_{\nu} T^{\mu\nu}_{cov} &= F^{\mu}_{\phantom{\mu}\nu}J_{cov}^{\nu} 
+c_g\ \epsilon^{\mu\nu} \nabla_\nu R.\\
\end{split}
\end{equation}

\subsection{Four dimensional theories}
\label{S:4danomalies}

We now turn to four-dimensional theories. The anomaly polynomial 6-form of an arbitrary
4-dimensional theory can be written in the form
\begin{equation}\label{E:anomP4d}
\mathcal{P}_{anom}[F,\mathfrak{R}] = c_{_A} F\wedge F\wedge F - 8\pi^2 c_m F\wedge p_{_1}(\mathfrak{R}) = c_{_A} F\wedge F\wedge F 
+ c_m F\wedge \text{tr}(\mathfrak{R} \wedge \mathfrak{R}) ,
\end{equation}
where $ c_{_A}$ is the $U(1)^3$ triangle anomaly coefficient and $c_m$ is the mixed 
$U(1)$-gravitational anomaly coefficient. Given a theory with chiral fermions, these coefficients can be calculated via
\begin{equation}
\begin{split}
c_{_A} &= -\frac{2\pi}{3!(2\pi)^3}\sum_{i=species} \chi_i(q_i^3)_{1/2} , \\
c_m &= -\frac{2\pi}{4!(8\pi^2)(2\pi)}\sum_{i=species} \chi_i (q_i)_{1/2} \,.
\end{split}
\end{equation}

The Chern-Simons form corresponding to the anomaly polynomial above is
\begin{equation}
I^{CS}_5 = A \wedge \brk{c_{_A} F\wedge F +(1-\alpha) c_m \text{tr}(\mathfrak{R} \wedge \mathfrak{R})}+ \alpha c_m F\wedge j_{CS}  . 
\end{equation}
where $j_{CS}$ is the gravitional Chern-Simons 3-form defined in \eqref{eq:gravCS} and 
$\alpha$ is a parameter which determines how the mixed anomaly is shared between the $U(1)$ and the 
gravitational transformations. It corresponds  to a gauge and diffeomorphism non-invariant contact term in the consistent generating function, $W_{\alpha} =-\int \alpha c_m A \wedge j_{CS}$. At $\alpha=1$ the mixed anomaly is completely associated with diffeomorphism transformations whereas for $\alpha=0$ the mixed anomaly is completely associated with $U(1)$ gauge transformations. This becomes clear if one works out the 
gauge and diffeomorphism transformation of the Chern-Simons form
\begin{equation}
\label{E:deltaI5}
\begin{split}
\delta_\lambda \int_{\mathcal{M}_5}I^{CS}_5 
&= \int_{\partial\mathcal{M}_5}  \Lambda\brk{c_{_A} F\wedge F +(1-\alpha) c_m \text{tr}(\mathfrak{R} \wedge \mathfrak{R})}+
\int_{\partial\mathcal{M}_5} \alpha\ c_m F\wedge  d\Gamma^{\lambda}{}_\nu (\partial_{\lambda}\xi^{\nu})  \,.
\end{split}
\end{equation}
Comparing \eqref{E:deltaI5} with \eqref{E:deltaCS} we get 
\begin{equation}
\begin{split}
\mathcal{J} &= \frac{1}{4}\epsilon^{\kappa\sigma\alpha\beta}\brk{   c_{_A}F_{\kappa\sigma}F_{\alpha\beta}
  +(1-\alpha) c_m R^\nu{}_{\lambda\kappa\sigma} R^\lambda{}_{\nu\alpha\beta} }, \\
\mathcal{T}^\lambda{}_\nu &= \alpha\ c_m \frac{1}{2}\epsilon^{\kappa\sigma\alpha\beta}
F_{\kappa\sigma}\partial_\alpha \Gamma^\lambda{}_{\nu\beta} ,  \\
\end{split}
\end{equation}
so that the consistent anomaly equations \eqref{eq:AnomCons} become
\begin{equation}
\begin{split}
\nabla_{\mu} J^{\mu}_{cons} &= \frac{1}{4}\epsilon^{\kappa\sigma\alpha\beta}\brk{   c_{_A}F_{\kappa\sigma}F_{\alpha\beta}
  +(1-\alpha) c_m R^\nu{}_{\lambda\kappa\sigma} R^\lambda{}_{\nu\alpha\beta} }, \\
\nabla_{\nu} T^{\mu\nu}_{cons} &= F^{\mu}{}_{\nu}J_{cons}^{\nu} 
-\frac{1}{4}\epsilon^{\kappa\sigma\alpha\beta}\brk{   c_{_A}F_{\kappa\sigma}F_{\alpha\beta}
  +(1-\alpha) c_m R^\nu{}_{\lambda\kappa\sigma} R^\lambda{}_{\nu\alpha\beta} } A^\mu\\
&\qquad - \alpha\ c_m \ g^{\mu\nu}\frac{1}{\sqrt{-g}}\ \partial_\lambda \brk{\sqrt{-g}\ \frac{1}{2}\epsilon^{\kappa\sigma\alpha\beta}
F_{\kappa\sigma}\partial_\alpha \Gamma^\lambda{}_{\nu\beta} },\\
\end{split}
\end{equation}

The variation of the Chern-Simons term with respect to sources is given by 
\begin{equation}
\begin{split}
\delta_S \int_{\mathcal{M}_5}I^{CS}_5
&= \int_{\mathcal{M}_5}  \Bigl\{\delta A \wedge \brk{3 c_{_A} F\wedge F + c_m \text{tr}(\mathfrak{R} \wedge \mathfrak{R})} 
+ 2 c_m \text{tr}(\delta\Gamma \wedge \mathfrak{R})\wedge F \Bigr\} \\
&\qquad + \int_{\partial\mathcal{M}_5}   \delta A \wedge \Bigl\{2 c_{_A} A\wedge F 
+ \alpha\ c_m j_{CS}\Bigr\} \\
&\qquad + \int_{\partial\mathcal{M}_5} \delta\Gamma^\nu{}_\lambda \wedge \Bigl\{\alpha\ c_m F\wedge\Gamma^\lambda{}_\nu+2(1-\alpha) c_m A\wedge\mathfrak{R}^\lambda{}_\nu  \Bigr\},\\
\end{split}
\end{equation}
from which we can read off the covariant anomaly and the Bardeen-Zumino currents using \eqref{E:deltaSCSdef}
\begin{equation}
\label{E:4dBZterms}
\begin{split}
  \mathrm{J}^{\bot} &=   \frac{1}{4}\epsilon^{\kappa\sigma\alpha\beta}\brk{   3 c_{_A}F_{\kappa\sigma}F_{\alpha\beta}
  + c_m R^\nu{}_{\lambda\kappa\sigma} R^\lambda{}_{\nu\alpha\beta} },\\
 \mathrm{L}^{\bot\mu\nu} &= 2 c_m  \frac{1}{4}\epsilon^{\kappa\sigma\alpha\beta}F_{\kappa\sigma}R^{\mu\nu}{}_{\alpha\beta},  \\
 P_{BZ}^\mu &= 2 c_{_A}\frac{1}{2}\epsilon^{\alpha\beta\mu\nu}F_{\alpha\beta}A_{\nu}+ \alpha\ c_m j^\mu_{CS},\\
  j^\mu_{CS}&\equiv \epsilon^{\mu\nu\kappa\sigma}\brk{
  \Gamma^\lambda{}_{\rho\nu}\partial_\kappa\Gamma^\rho{}_{\lambda\sigma} 
  +\frac{2}{3}\Gamma^\lambda{}_{\alpha\nu}\Gamma^\alpha{}_{\rho\kappa}\Gamma^\rho{}_{\lambda\sigma} },\\
X^{\mu\lambda}{}_\nu &=
\alpha\ c_m  \frac{1}{2}\brk{\epsilon^{\mu\rho\kappa\sigma}\Gamma^\lambda{}_{\nu\rho}+
\epsilon^{\lambda\rho\kappa\sigma}\Gamma^\mu{}_{\nu\rho}}F_{\kappa\sigma}\\
&\qquad+2(1-\alpha) c_m \frac{1}{2}\brk{ \epsilon^{\mu\rho\kappa\sigma} R^\lambda{}_{\nu\kappa\sigma}
+\epsilon^{\lambda\rho\kappa\sigma} R^\mu{}_{\nu\kappa\sigma}}A_\rho,\\
P^{\mu\nu}_{BZ} &=
-\frac{1}{2} \nabla_{\lambda}\prn{ 
X^{\lambda\mu\nu}+X^{\lambda\nu\mu}-X^{\mu\nu\lambda}}\,.
\end{split}
\end{equation}
The covariant anomaly equations \eqref{eq:Anomcov} are now given by
\begin{equation}\label{E:nonC4d}
\begin{split}
\nabla_{\mu} J^{\mu}_{cov} &=  \frac{1}{4}\epsilon^{\kappa\sigma\alpha\beta}\brk{   3 c_{_A}F_{\kappa\sigma}F_{\alpha\beta}
  + c_m R^\nu{}_{\lambda\kappa\sigma} R^\lambda{}_{\nu\alpha\beta} },\\
\nabla_{\nu} T^{\mu\nu}_{cov} &= F^{\mu}_{\phantom{\mu}\nu}J_{cov}^{\nu} 
+2 c_m\  \nabla_\nu \brk{\frac{1}{4}\epsilon^{\kappa\sigma\alpha\beta}F_{\kappa\sigma}R^{\mu\nu}{}_{\alpha\beta} }.\\
\end{split}
\end{equation}
Note that the final covariant anomaly equations do not depend on the parameter $\alpha$, which should not surprise us as $\alpha$ corresponds to a local counterterm. We see that the mixed anomaly is symmetrically shared between the $U(1)$ and gravitational currents.

\section{The Euclidean generating functional}
\label{S:generating}
In this section we review the method by which the Euclidean partition function for theories in thermodynamic equilibrium may be constructed in terms of a functional of the sources. Most of the material contained here is a shortened version of \cite{Jensen:2012jh,Banerjee:2012iz} to which we refer the reader for more details.

\subsection{The generating function}
\label{S:genReview}
We begin by reviewing the arguments presented in~\cite{Jensen:2012jh} used to construct the generating function of a thermodynamic theory.  Consider a Lorentz-invariant quantum field theory in $d$ space-time dimensions at nonzero temperature $T$ in flat space. 
Generically, the real-space Euclidean correlators of this theory will decay exponentially at large distances implying that the screening lengths $\xi_i$ of the theory,
\begin{equation}
\label{E:screening}
\langle \mathcal{O}(\tau,\bf{x})\mathcal{O}(\tau,0)\rangle \sim \exp \left(-|\bf{x}| / \xi \right),
\end{equation}
are finite. The assumption that the correlation length is finite implies that, for example, we are not at a critical point of the theory, or that there are no unscreened long-range forces as in QED. Instead of screening lengths, one may consider screening masses which are given by the location of the poles of the momentum-space zero-frequency Euclidean correlation functions along the imaginary momentum axis. The screening lengths are inversely proportional to the screening masses.

From \eqref{E:screening} it follows that the zero-frequency correlation functions of the theory are analytic at zero spatial momentum. We then define {truncated correlation functions} by Taylor-expanding all zero-frequency correlation functions to $m^{th}$ order around zero spatial momentum. The position-space truncated correlators may then be obtained by varying a generating function: an integral of a local functional of background fields, which we denote $W_m$. In a theory with local gauge and diffeomorphism invariance, $W_m$ may be written as an integral of gauge and differmorphism invariant (up to boundary terms) scalars. In the presence of anomalies, appropriate local (but Lorentz breaking) terms may be introduced into $W_m$ in order to account for its anomalous variation as described in section \ref{S:AP}.

Due to the non-analytic behavior of Euclidean correlators in the complex momentum plane, one can expect the derivative expansion to have a finite radius of convergence at best. Nevertheless, one may always carry out a formal expansion of the Euclidean generating function around the origin to arbitrary order. A resummation of this series will presumably uncover the poles or branch cuts of $W$. Terms which are non-analytic at zero momentum, e.g., $e^{-1/k}$, will not be accounted for in our construction and must be included by hand when studying properties of the theory which go beyond the derivative expansion. We will discuss the role that such terms play in our analysis in sections~\ref{S:2d} and~\ref{S:4d}.

Put differently, were we able to resum the derivative expansion and uncover all non-analytic contributions, we would have obtained the equilbirum partition function. By equilibrium partition function we mean the Euclidean partition function evaluated on a background with time independent sources. In practice this resummation is unfeasible, but even in configurations where the derivative expansion breaks down one may still extract information about correlation functions without performing the resummation. We follow this method in sections~\ref{S:2d} and~\ref{S:4d}.

To make the construction of $W_m$ explicit, consider a Euclidean theory defined on a manifold $\mathcal{M}_d=\Sone\times_f \mathcal{M}_{d-1}$, where the $\mathbb{S}^1$ is the time circle which may be fibered non-trivially over the spatial manifold $\mathcal{M}_{d-1}$. Let $t$ denote time and $V = \partial_t$, i.e., $V^{\mu}$ is the vector whose integral curves give the time circle. In thermal equilibrium we demand that $\pounds_V=0$: the Lie derivative of all sources with respect to $V$ vanish. Consequentially $V$ is a Killing vector. The generating function $W_m$ may be constructed of all possible gauge and diffeomorphism invariant combinations of the metric, gauge field, Killing vector $V$ and non-local quantities which may be constructed using $V$, namely the inverse length of the time circle $T^{-1}=\int_0^{\beta} \sqrt{-V^2} d\tau$, which we identify with the temperature, and the Polyakov loop $P_A = \exp(\int_0^{\beta} A_{\mu}V^{\mu} d\tau)$. Here $\tau = i t$ denotes Euclidean time. We find it convenient to replace $V^{\mu}$ with the fluid velocity $u^{\mu} = V^{\mu}/\sqrt{-V^2}$ and the Polyakov loop with a chemical potential $\mu = T \ln P_A$. 

Our conventions are such that varying $W$ with respect to the sources as in \eqref{E:TJCons} will result in real one-point functions in Lorentzian signature. That is, we work with a Wick rotated generating function. If we denote the partition function of the theory by $Z $ then we define $W = - i \ln Z$.  By construction, all the covariant $n$ point functions transform covariantly when Wick rotating to or from Euclidean time.

Let us work out a simple example in detail. We wish to construct the generating function of a parity preserving theory to first order in the derivative expansion. To do so, we must enumerate all possible scalar expressions which have zero or one derivatives acting on gauge and diffeomorphism invariant combinations of $T$, $\mu$, $g_{\mu\nu}$ and $A_{\mu}$. Consider the identities,
\begin{equation}
\label{E:derivatives}
\nabla_{\mu}u_{\nu}=-u_{\mu}a_{\nu} + \Omega_{\mu\nu}, \qquad \nabla_{\mu}T = -T a_{\mu}, \qquad \nabla_{\mu}\mu = -\mu a_{\mu}+E_{\mu},
\end{equation}
where we have defined
\begin{equation}
a_{\mu} = u^{\nu}\nabla_{\nu}u_{\mu},\qquad \Omega^{\mu\nu}=\frac{1}{2}\Delta^{\mu\rho}\Delta^{\nu\sigma}(\nabla_{\rho}u_{\sigma}-\nabla_{\sigma}u_{\rho}), \qquad \Delta^{\mu\nu}=g^{\mu\nu}+u^{\mu}u^{\nu}, \qquad E_{\mu}=F_{\mu\nu}u^{\nu}.
\end{equation}
These identities follow from the fact that $\pounds_V=0$ when acting on sources. The vector $a^{\mu}$ is the acceleration vector, $\Omega^{\mu\nu}$ is the vorticity tensor, and $E^{\mu}$ is the electric field. Since $a^{\mu}$, $\Omega^{\mu\nu},$ and $E^{\mu}$ are all transverse to $u^{\mu}$, there are no local gauge-invariant scalars with one derivative. Thus,
\begin{equation}
W_1 = \int d^{d}x\sqrt{-g} P(T,\mu)\,.
\end{equation}
We refer the reader to \cite{Jensen:2012jh} for more details.

In~\cite{Banerjee:2012iz} an alternate method was used to obtain the generating function. By choosing coordinates such that $V=\partial_t$, the $d$-dimensional metric may be written as 
\begin{equation}
\label{E:theirg}
	g = -e^{2\sigma(x)}(dt+a_i(x) dx^i)^2 + g^{(d-1)}_{ij}(x)dx^idx^j,
\end{equation}
where $g^{(d-1)}_{ij}$ is the metric of the spatial manifold $\mathcal{M}_{d-1}$. The authors of~\cite{Banerjee:2012iz} constructed the generating function on $\mathcal{M}_{d-1}$ by Wick-rotating to Euclidean signature and then dimensionally reducing on the time circle. The dimensionally reduced generating function may be constructed out of the the dimensionally reduced metric, the gauge field, and the Kaluza-Klein (KK) photon $a_i$. For instance, as discussed in section 2.2 of~\cite{Banerjee:2012iz}, in order to preserve gauge invariance, KK invariance, and preserve the form \eqref{E:theirg} the spatial components of the gauge field, $A_i$ must appear in the combination,\footnote{In the language of \cite{Banerjee:2012iz} we have $A^{\mu}_{ours} = \mathcal{A}^{\mu}_{theirs}$, $T_{ours} = \exp(-\sigma)/\beta\Big|_{theirs}$, $\mu_{ours} =  \exp(-\sigma) A_0 \Big|_{theirs}$.}
\begin{equation}
	\mathcal{A}_i = A_i - a_i A_0.
\end{equation}	
The main difference between the methods of \cite{Jensen:2012jh} and those of \cite{Banerjee:2012iz} are in the symmetries which are made manifest. In  \cite{Jensen:2012jh} the full $d$-dimensional diffeomorphism and gauge invariances are apparent, while in~\cite{Banerjee:2012iz} the spatial diffeomorphisms and the symmetry generated by $V$ are clearly visible.

\subsection{Chern-Simons contributions to the generating function}
\label{S:CS}
As we have alluded to in the previous section, in addition to scalars built out of gauge-invariant tensors, we may add to the generating function terms which are gauge and diffeomorphism-invariant up to boundary terms. We presently classify all such terms, as they will play a crucial role in the sections to come. Given a conserved, gauge-invariant current $X^{\mu}$ which is transverse to $u^{\mu}$ we may construct
\begin{equation}
\label{E:WCS}
W_X = \int d^{d}x \sqrt{-g}\left(  c_{1} A_{\mu} X^{\mu} +c_{2} X^0\right),
\end{equation}
where the $c_{i}$ are constants and we choose coordinates so that $V = \partial_t$. (If $X^{\mu}$ is parallel to $u^{\mu}$ then both terms in \eqref{E:WCS} reduce to scalar expressions.) The first term on the right hand side of \eqref{E:WCS} is gauge invariant up to boundary terms. The second term is diffeomorphism-invariant up to boundary terms due to the fact that $V$ is Killing. Since $A_0$ is odd under CPT, either $c_1$ or $c_2$ should vanish in a CPT preserving theory.

In the approach of~\cite{Banerjee:2012iz}~\eqref{E:WCS} may be rewritten in the form
\begin{equation}
\label{E:WXspatial}
	W_X = \beta\int d^{d-1}x\, \sqrt{g^{(d-1)}}\left( c_{1}\mathcal{A}_i\mathcal{X}^i  - c_{2}a_i \mathcal{X}^i \right),
\end{equation}
where we have defined the Kaluza-Klein invariant covector $\mathcal{X}_i=e^{\sigma}\left( X_i-a_i X_0 \right)$. Since $X_{\mu}u^{\mu} = 0$ by definition then $\mathcal{X}_i$ reduces to $\mathcal{X}_i = e^{\sigma}X_i$. In this notation the spatial indices $i,j$ are raised and lowered with the spatial metric $g^{(d-1)}_{ij}$ defined in \eqref{E:theirg} and its inverse. 

If we denote the Hodge star operator on $\mathcal{M}_{d-1}$ by $\hs$ and treat $\mathcal{A}_i$ and $\mathcal{X}_i$ as one-forms then \eqref{E:WXspatial} can be recast as
\begin{equation}
\label{E:WXform}
W_X = \beta \int c_1 \mathcal{A}\wedge \hs \mathcal{X} -c_2 a \wedge \hs \mathcal{X}\,.
\end{equation}
We would like to argue that $\hs \mathcal{X}$ must either be a product of Chern classes of $\mathcal{A}$ and $a$ ($d\mathcal{A}$ and $da$) and Pontryagin classes of the Riemann tensor, or that \eqref{E:WXform} reduces to the gauge-invariant scalars and pseudo scalars described in section \ref{S:genReview}.

Conservation of $X^{\mu}$ implies that $\mathcal{X}$ is conserved, $d\hs  \mathcal{X} = 0$. On a spatial manifold with trivial topology there are no non-trivial cohomology groups and every closed form is exact. Thus, $\hs \mathcal{X}=d\mathcal{Y}$. If $\mathcal{Y}$ is gauge-invariant then after integrating \eqref{E:WXform} by parts we obtain
\begin{equation}
W_X = \beta \int c_1 d\mathcal{A}\wedge \mathcal{Y}  - c_2 da\wedge \mathcal{Y}\,,
\end{equation}
implying that \eqref{E:WXspatial} may be rewritten in terms of the gauge-invariant scalar contributions to $W$ described in Section~\ref{S:genReview}. If $\mathcal{Y}$ is not gauge invariant, then $d\mathcal{Y}$ must be gauge invariant in which case the gauge variation of $\mathcal{Y}$, $\delta_{\lambda}\mathcal{Y}$ must be exact on a topologically trivial manifold, $\delta_{\lambda}\mathcal{Y} = d\mathcal{Z}$. Thus, $\mathcal{Y}$ must a linear combination of $d-3$-dimensional Chern-Simons forms implying that $\hs \mathcal{X} = d\mathcal{Y}$ is a product of Chern classes of the $U(1)$ gauge fields and Pontryagin classes of the spatial metric.  Put differently,  the terms in~\eqref{E:WXspatial} are genuine Chern-Simons terms on $\mathcal{M}_{d-1}$.

\section{Thermodynamics in two dimensions}
\label{S:2d}
In the following sections we will construct the most general Euclidean generating functional for a two-dimensional Lorentz-invariant theory to second order in derivatives. In section~\ref{S:generating2d} we construct the most general generating function to second order in the derivative expansion and derive the appropriate constitutive relations.

We place our theory on a Euclidean cone with metric
\begin{equation}
\label{E:cone}
	ds^2 = dr^2 + r^2 d\tau^2,
\end{equation}
where $\tau$ has periodicity $\tau\sim\tau+2\pi\delta$ and it parameterizes the Euclidean time circle. The parameter $\delta$ is related to the deficit angle of the cone, $2\pi (1-\delta)$, thereby interpolating between a cylinder ($\delta \to 0$) associated with a thermal state of the theory, and $\mathbb{R}^2$ ($\delta\to 1$) associated with the Euclidean vacuum. 
The expectation value of the stress tensor in the Euclidean vacuum vanishes in a conformal theory, and is proportional to the metric in more general theories. This property of the stress tensor imposes restrictions on the thermal theory, as we now show.\footnote{The $\delta \to 0$ limit seems somewhat singular in the coordinate system~\eqref{E:cone}. This can be remedied by going to a coordinate system defined such that $r = \frac{R}{\delta} e^{\frac{\delta \rho}{R}}$ and $\theta = \phi\,\delta$ which is valid for any non zero $\delta$ where the line element takes the form $ds^2 = e^{\frac{2 \delta \rho}{R}} \left( d\rho^2 + R^2 d\phi^2 \right)$ whose $\delta \to 0$ limit gives us the cylinder.}

\subsection{The generating functional and constitutive relations}
\label{S:generating2d}
Consider a two dimensional relativistic theory on a manifold $\mathcal{M}$ with metric $g_{\mu\nu}$ in the presence of a gauge field $A_{\mu}$. We wish to construct the thermal partition function for this theory up to second order in the derivative expansion following the algorithm presented in Section~\ref{S:generating}. To this end, we need to enumerate all possible scalars, pseudo scalars and conserved currents to second order in the derivative expansion assuming that the Lie derivative of the metric and gauge field vanish for a timelike vector $V^{\mu}$. 
We have listed all gauge-invariant scalars, pseudo scalars and conserved currents to second order in the derivative expansion in Table~\ref{T:all}. The alert reader may note that no components of the Riemann tensor appear in Table~\ref{T:all}. This is because of the fact that, in a two-dimensional time-independent background, the Riemann tensor is given by
\begin{equation}
\label{E:2dRiemann}
R_{\mu\nu\rho\sigma} = \left( \frac{s_5}{T}-s_1\right)(g_{\mu\rho}g_{\nu\sigma}-g_{\mu\sigma}g_{\nu\rho})\,,
\end{equation}
where $s_1$ and $s_5$ are defined in Table~\ref{T:all}, and so the Riemann tensor is not independent. Thus, to second order in derivatives the most general generating functional takes the form 
\begin{equation}
\label{E:Wdef}
	W_2 = \int d^2x\sqrt{-g} \left(P(T,\mu) + \tilde{c}_{2d} \beta^{-1} X^0 + \tilde{s}_1  \tilde{\alpha}_1(T,\mu) + \sum_{i=1}^3 s_i \alpha_i(T,\mu)  \right)\,.
\end{equation}
We have not included the scalars $s_4$ and $s_5$ in \eqref{E:Wdef} since they are total derivatives which, after partial integration, contribute to $\alpha_1$ and $\alpha_3$. Similarly, there exists a linear combination of $\tilde{s}_1$ and $\tilde{s}_2$ which vanishes after integrating by parts. Thus, we have omitted a term involving $\tilde{s}_2$. The coefficient of the Chern-Simons term $X^0$ has been denoted $\tilde{c}_{2d}$ to distinguish it from the coefficient of the Chern-Simons term of a four dimensional theory to be introduced in the next section.
\begin{table}[ht]
\begin{center}
\begin{tabular}{| l | l | l |}
\hline
	1st order & 2nd order & conserved currents \\
\hline
		$\tilde{ s}_{1} = \nabla_{\mu}\tilde{u}^{\mu}$    		& 	$s_{1} = a_{\mu}a^{\mu} $ & $X^{\mu}=T\tilde{u}^{\mu}$	\\ 
	        	$\tilde{s}_{2} =\epsilon^{\mu\nu}F_{\mu\nu}$ 		&  	$s_{2} = F_{\mu\nu}F^{\mu\nu}$  & \\
			& $s_{3} = \nabla_{\mu}T \nabla^{\mu} \frac{\mu}{T}$  & \\
			& $s_{4} = \nabla_{\mu}\nabla^{\mu} \frac{\mu}{T} $ & \\
			& $s_{5} = \nabla_{\mu}\nabla^{\mu} T$ &  \\										
\hline
\end{tabular}
\caption{ \label{T:all}  A list of all gauge-invariant scalars and conserved currents formed out of the background fields and $V$ with at most two derivatives. The quantities $u^{\mu}$, $T$, $\mu$ and $a^{\mu}$ were defined in Section~\protect\ref{S:genReview}. 
The transverse vector $\tilde{u}^{\mu}$ is defined through $\tilde{u}^{\mu} = \epsilon^{\mu\nu}u_{\nu}$.}
\end{center}
\end{table}

If the theory has a $U(1)$ or a gravitational anomaly then we must add to the generating function extra terms which break gauge and diffeomorphism invariance such that the resulting current and stress tensor are anomalous. We find that the appropriate terms we need to add to the generating function are given by
\begin{equation}
\label{E:2dWA}
	W_A = \int d^2x \sqrt{-g} \left(-{c_s} \mu A_{\mu}\tilde{u}^{\mu} - c_g u_\alpha u^\beta \epsilon^{\mu\nu} \partial_{\mu} \Gamma^{\alpha}_{\phantom{\alpha}\beta\nu} \right)\,.
\end{equation}
In \cite{Valle:2012em} the second expression in~\eqref{E:2dWA} has been obtained by integrating the equations of motion. The covariant energy momentum tensor ${T}_{cov}^{\mu\nu}$ and current ${J}_{cov}^{\mu}$ are given by~\eqref{E:covJT} with $W=W_2+W_A$, and one can check that the resulting (non)conservation equations are given by~\eqref{E:2dcovCons},
\begin{equation}
	\nabla_{\mu}{J}_{cov}^{\mu}=c_s \epsilon^{\mu\nu}F_{\mu\nu}, 
	\qquad
	\nabla_{\mu}{T}_{cov}^{\mu\nu}=F^{\nu}{}_{\rho}{J}^{\rho}_{cov}+c_g \epsilon^{\nu\rho}\partial_{\rho}R.
\end{equation}
An explicit computation gives the one-point functions of the covariant current and stress tensor to be
\begin{align}
	J_{cov}^{\mu} = R u^{\mu} + \tilde{R} \tilde{u}^{\mu},
	\qquad
	T_{cov}^{\mu\nu} = \mathcal{E} u^{\mu}u^{\nu} + \mathcal{P} \tilde{u}^{\mu}\tilde{u}^{\nu} + \theta \left( u^{\mu}\tilde{u}^{\nu} + u^{\nu}\tilde{u}^{\mu}\right)
\end{align}
where we define $\tilde{u}^{\mu}$ was defined in the caption of table~\ref{T:all} and
{\small
\begin{align}
\notag
	R =& \frac{\partial P}{\partial \mu}
		+ \ddm{\tilde{\alpha}_1} \tilde{s}_1
		+\left(\ddm{\alpha_1} - \mu \ddm{\alpha_3} - T \ddT{\alpha_3} \right)s_1
		-\ddm{\alpha_2}s_2
		+4 \left( 2 \alpha_2 + \mu \ddm{\alpha_2} + T \ddm{\alpha_2}\right)s_3
		+4 T \alpha_2 s_4
		-\frac{\alpha_3}{T} s_5\, ,
	\\
\notag
	\tilde{R} =&  - 2 c_s \mu\, , 
	\\
\notag
	\mathcal{E} =& -P + \mu \frac{\partial P}{\partial \mu} + T \frac{\partial P}{\partial T} 
		-\frac{1}{2} \ddm{\tilde{\alpha}_1} \tilde{s}_2
		+\left(\alpha_1 - \mu \ddm{\alpha_1} - T \ddT{\alpha_1} \right) s_1
		+\left(\alpha_2 +\mu \ddm{\alpha_2} + T \ddT{\alpha_2} + \frac{1}{2} \ddm{\alpha_3} \right) s_2 
	\\
\label{E:general}
		&-\left(\alpha_3+2\ddm{\alpha_1}\right)s_3 
		- T \alpha_3 s_4 
		- \frac{2 \alpha_1}{T} s_5 ,
	\\
\notag
	\mathcal{P} =& P - \alpha_1 s_1 - \alpha_2 s_2 - \alpha_3 s_3 ,
	\\
\notag
	\theta =& \tilde{c}_{2d} T^2 - c_s \mu^2 + \frac{2c_g}{T} s_5\,.
\end{align}
}

\subsection{Conformal theories}
\label{S:2dcft}
In a conformal theory and in the absence of anomalies, the generating function must be invariant under Weyl rescalings of the metric. In terms of the parameters in \eqref{E:Wdef} we find that this implies that $P = T^2 p_0(\mu/T)$, $\tilde{\alpha}_1 = 0$,\footnote{%
In practice, if we set $\tilde{\alpha}_1 = \tilde{\alpha}_1(T)$ then the expression $\sqrt{-g}\tilde{\alpha}_1 \tilde{s}_1$ becomes a total derivative and does not contribute to the energy momentum tensor and current. Similarly, had we used $\alpha_2(\mu/T) \tilde{s}_2$ in place of $\alpha_1(T)\tilde{s}_1$ we would have found that it contributes only to a boundary term.} 
$\alpha_1 = \alpha_3=0$ and that $\alpha_2 = a_2(\mu/T)/T^2$. In addition to Weyl-invariance, conformal symmetry enlarges the $U(1)$ gauge symmetry to a doublet of current algebras for a left-moving and right-moving current. This doublet gives additional constraints on the generating functional, owing to the independent conservation laws for the chiral currents. For instance,  the chiral anomaly fixes the dynamics of $U(1)$ currents when  in the high temperature limit, when the chemical potential vanishes, or when the $U(1)$ is non-compact~\cite{Kraus:2006wn,Jensen:2010em}. (Compact $U(1)$s may be more interesting due to non-perturbative effects, like the twisting of winding modes in the presence of a chemical potential.) In the remainder of this section we will ignore the effects of the doublet of current algebras and leave them for future inquiry.

In the absence of a $U(1)$ current we would have $a_2=0$ so that all first and second order terms in the derivative expansion are excluded. In fact, conformal invariance precludes us from writing any higher derivative correction to the generating function. Indeed, suppose that there exists a scalar expression which contributes to the generating function. Such a scalar must be constructed out of the Weyl tensor $W_{\mu\nu\rho\sigma}$, the velocity field $u_{\mu}$, the temperature $T$ and their gradients. Since gradients of Weyl covariant quantities generate non Weyl covariant expressions it is convenient to trade the gradient $\nabla_{\mu}$ with a Weyl covariant one $\nabla^W_{\mu}$ defined in \cite{Loganayagam:2008is}. Since $\nabla^W_{\mu} u_{\nu} = 0$, $\nabla^W_{\mu} T  = 0$ and the Weyl tensor identically vanishes in two dimensions, there are no Weyl-covariant tensors with one or more derivatives. A moment's thought also shows that we may not add any terms non-analytic in derivatives either. Furthermore, the analysis of section~\ref{S:CS} implies that there are no conserved
currents other than $X^{\mu}$ defined in table~\ref{T:all}. It then follows that the equilibrium partition function of an uncharged conformal fluid contains only the pressure and anomaly terms.

Often, the Weyl symmetry of a conformal theory will be anomalous. This anomaly manifests itself as a non-vanishing trace of the stress tensor
\begin{equation}
\label{E:WeylAnom}
	\left( {T}_{cov}\right)_{\mu}^{\mu} = c_{w} R\,,
\end{equation}
where $c_{w}$ is proportional to the central charge as defined in \eqref{eq:cwCFT} and $R$ is the Ricci scalar.
The Weyl anomaly \eqref{E:WeylAnom} can be generated by allowing for a non-Weyl covariant contribution to the generating function of the form $\alpha_1=c_{w}$. Thus, in the presence of gravitational, gauge and Weyl anomalies, the generating functional of a conformal theory will take the form
\begin{equation}
	W_2 = \int d^2x \sqrt{-g} \left( T^2p_0\left({\mu}/{T}\right) +\tilde{c}_{2d} \beta^{-1} X^0  + \frac{a_2(\mu/T) s_2}{T^2}  + c_{w} s_1
		\right) + W_A\, ,
\end{equation}
where $W_A$ was given in~\eqref{E:2dWA}. Using \eqref{E:general} we can write the resulting energy momentum tensor and current in the form
\begin{align}
\begin{split}
\label{E:conformal}
	R =& \frac{\partial P}{\partial \mu}
		-\frac{a_2'}{T^3} s_2
		+\frac{4 a_2}{T} s_4,
	\\
	\tilde{R} =& - 2 c_s \mu ,
	\\
	\mathcal{E} =& p_0 T^2
		+c_{w} s_1
		-\frac{a_2}{T^2} s_2 
		- \frac{2 c_{w}}{T} s_5 ,
	\\
	\mathcal{P} =& p_0 T^2 - c_{w} s_1 - \frac{a_2}{T^2} s_2 , \\
	\theta =& \tilde{c}_{2d} T^2 - c_s \mu^2 + \frac{2c_g}{T} s_5\,.
\end{split}
\end{align}
Note that 
\begin{equation}
\label{E:2dderivatives}
	\nabla_{\mu}\nabla^{\mu} \frac{\mu}{T} = \frac{1}{\sqrt{-g}} \partial_{\mu} \sqrt{-g} g^{\mu\nu}\partial_{\nu} \frac{\mu}{T}
\end{equation}
transforms homogeneously under Weyl transformations with weight $+2$ in two dimensions. As explained previously, in the absence of a $U(1)$ current the relation~\eqref{E:conformal} is exact. We note that relation \eqref{E:conformal} is exact also in the absence of an external electric field wherein gradients of the chemical potential are proportional to the acceleration such that $\nabla_{\alpha} \left(\mu/T\right) = 0$ (and then $s_2 = s_4 = 0$).

Let us place our theory on a cone with line element
\begin{equation}
	ds^2 =dr^2+ r^2 d\tau^2,
\end{equation}
with $\tau \sim \tau+2\pi\delta$ so that $2\pi(1-\delta)$ is the deficit angle of the cone. The limit $\delta \to 1$ corresponds to Euclidean space and $\delta \to 0$ corresponds to the cylinder. Treating the angular coordinate as the Euclidean time direction, the temperature of the theory on the cone is given by
\begin{equation}
	T^{-1} = 2\pi \delta r\,.
\end{equation}
Thus, we have, for example, $\nabla_{\mu} T^{-1} \nabla^{\mu}T^{-1} = 4 \pi^2 \delta^2$, and the gradient expansion on the cone breaks down. Despite the breakdown of the derivative expansion, the theory on the cone is in a type of time-independent equilibrium state. In this time-independent state, we continue to formally define the temperature $T$ as the inverse length of the time circle. As we have just argued, the generating function $W$ is the most general Euclidean generating function on can write down for a two dimensional conformal theory. Therefore, its variation will correctly capture the expectation value and current at least far from the tip of the cone.

For an uncharged conformal theory on the cone,
\begin{equation}
\label{E:Tmn2dcft}
	{T}_{cov}^{\mu\nu} = (p_0-4\pi^2 c_{w} \delta^2)T^2(2 u^{\mu}u^{\nu}+g^{\mu\nu})
		+(\tilde{c}_{2d}+8\pi^2\delta^2c_g)T^2(u^{\mu}\tilde{u}^{\nu}+u^{\nu}\tilde{u}^{\mu})\,.
\end{equation}
In the limit where $\delta=1$ we expect to recover the energy momentum tensor in the Euclidean vacuum of a conformal theory on $\mathbb{R}^2$ which should vanish. Thus,
\begin{subequations}
\label{E:result2dcft}
\begin{align}
\label{E:p0val}
	p_0 &= 4 \pi^2 c_w\,,\\
\label{E:c2val}
	 \tilde{c}_{2d} &= -8 \pi^2 c_g\,.
\end{align}
\end{subequations}
We note in passing that instead of placing the theory on a cone, one could have placed the theory on a punctured plane $\mathbb{R}^2/\{0,0\}$ and carried out the entire calculation by setting $\delta=1$ from the outset. One would then require that after adding the missing point at the origin, the stress tensor~\eqref{E:result2dcft} should coincide with the stress tensor of the Euclidean vacuum, i.e., it should vanish. We make some further comments about this alternate viewpoint in section~\ref{S:Dlimits}.

Once there is a $U(1)$ current, we need to account for higher derivative terms in the generating functional and one might worry that these will spoil \eqref{E:c2val}. However, we are interested only in those terms which do not vanish when we set the external gauge field to zero so the only worrisome terms which we need to account for are those that contain gradients of the chemical potential. As noted previously, in the absence of an external electric field $\nabla_{\alpha}(\mu/T) = 0$. Thus, \eqref{E:p0val} remains valid to leading order in the chemical potential, and \eqref{E:c2val} does not get corrected
\[
	p_0 = 4 \pi^2 c_w + \mathcal{O}(\mu) \,,
	\qquad
	\tilde{c}_{2d} = - 8 \pi^2 c_g\,.
\]

\subsection{Non-conformal theories}
\label{S:2dncft}
An important ingredient in obtaining \eqref{E:result2dcft} was the absence of higher derivative contributions to the generating function. Had there been higher derivative corrections to the generating function, the breakdown of the derivative expansion on the cone would have forced us to consider the possible contribution of any such corrections to the energy momentum tensor. We now turn to such considerations.

Using \eqref{E:general} to compute $T^{\tau r}_{cov}$  explicitly, we find
\begin{equation}
\label{E:2tr2d}
	T^{\tau r}_{cov} = i\frac{\tilde{c}_{2d} + 8 \pi^2 \delta^2 c_g}{4 \pi^2 r^3 \delta^2}\,.
\end{equation}
We would like to argue that $T^{\tau r}_{cov}$ does not receive corrections at any finite order in the derivative expansion.
To this end, consider the perturbed cone, given by
\begin{equation}
\label{E:perturbedcone}
	g = dr^2 + r^2 d\tau^2 + h_{\tau r}(r) drd\tau\,.
\end{equation}
That corrections to ${T}^{\tau r}_{cov}$ are absent amounts to the statement that there are no scalars or conserved currents which we can add to the generating function which are linear in $h_{\tau r}$. We will now prove such a statement. 

In the absence of external gauge fields, the tensors that may contribute to the Euclidean generating functional are built out of $u_{\mu}$, $T$, $\mu$, $R_{\mu\nu\rho\sigma}$, and derivatives thereof. In two dimensions the vorticity tensor vanishes so that the general expression~\eqref{E:derivatives} becomes
\begin{equation}
	\nabla_{\mu}T = - T a_{\mu}, \qquad 
	\nabla_{\mu} \frac{\mu}{T} = 0 \qquad
	\nabla_{\mu}u_{\nu}=-u_{\mu}a_{\nu}.
\end{equation}
The Riemann tensor in two dimensions takes the form~\eqref{E:2dRiemann},
and on the cone, to linear order in $h_{\tau r}$,
\begin{equation}
\nabla_{\mu}a_{\nu}=-u_{\mu}u_{\nu}a^2-a_{\mu}a_{\nu}.
\end{equation}
Thus, the most general scalar expression in a two dimensional thermodynamic theory may be constructed out of $a_{\mu}$, $u_{\mu}$, the metric $g_{\mu\nu}$ and the epsilon tensor $\epsilon_{\mu\nu}$ (and no derivatives). A quick computation then shows that one can not construct a scalar linear in $h_{\tau r}$.  Following the discussion in section \ref{S:CS} there are no conserved currents in two dimensions other than $X^{\mu}$ which will contribute non trivially to the generating function. 
Therefore, at any finite order in the derivative expansion there are no gauge or diffeomorphism invariant expressions which can contribute to $T^{\tau r}$ on the cone in the absence of sources. We also note that our argument precludes local non perturbative terms from contributing to $T^{\tau r}$ as well. For example, an expression of the form $e^{-\frac{T^n}{s}}$, with $T$ the temperature $s$ a scalar and $n$ some number, can not contribute to $T^{\tau r}$ since $s$ has no contribution linear in $h_{\tau r}$ when evaluated on the background \eqref{E:perturbedcone}.

At $\delta = 1$, $T^{\tau r}_{cov}$ should vanish by the translational invariance of the Euclidean vacuum. Since $T^{\tau r}_{cov}$ is given precisely by \eqref{E:2tr2d} then requiring it to vanish in the limit $\delta \to 1$ implies that
\[
	\tilde{c}_{2d} = - 8 \pi^2 c_g\,.
\]
This demonstrates our claim that \eqref{E:c2val} remains valid for non conformal theories.

\section{Four dimensional theories}
\label{S:4d}
We move our attention to four dimensional theories with anomalies. We start by constructing the gauge and diffeomorphism invariant generating function to first order in derivatives. As mentioned in Section~\ref{S:genReview}, there are no local gauge-invariant scalars with one derivative but there is one (and only one) CPT preserving Chern-Simons term of the type described in section \ref{S:CS} given by $a \wedge d\mathcal{A}$. The corresponding conserved current is
\begin{equation}
	X_1^{\mu}=T(B^{\mu}+\mu \omega^{\mu})\,,
\end{equation}
where the magnetic field $B^{\mu}$ and vorticity $\omega^{\mu}$ are defined by
\begin{equation}
B^{\mu}=\frac{1}{2}\epsilon^{\mu\nu\rho\sigma}u_{\nu}F_{\rho\sigma}, \qquad \omega^{\mu}=\epsilon^{\mu\nu\rho\sigma}u_{\nu}\nabla_{\rho}u_{\sigma}\,,
\end{equation}
In the absence of anomalies, the generating function to first order in derivatives is given by
\begin{equation}
\label{E:W14d}
	W_1 = \int d^4x \sqrt{-g} \left(P(T,\mu) + \tilde{c}_{4d} \beta^{-1} X_1^{0} \right).
\end{equation}
The alert reader will note that there exists an additional conserved current $X_2^{\mu} = T^2 \omega^{\mu}$ which arguably contributes to the generating function a term of the form $W_X = \int d^4x \sqrt{-g}  \tilde{c} A_{\mu}X^{\mu}_2$. However, the scalar $A_{\mu}X_2^{\mu}$ corresponds to the Chern-Simons term $\mathcal{A}\wedge da$ which is equivalent to $a \wedge d\mathcal{A}$ after integrating by parts. Indeed, $\int d^4x\sqrt{-g} \beta^{-1} X_1^0 = \int d^4x\sqrt{-g} A_{\mu}X_2^{\mu} $ up to boundary terms.

If the theory has $U(1)^3$ and mixed anomalies, the partition function has an anomalous variation given by (minus)~\eqref{E:deltaI5}. Such an anomalous variation is generated by
\begin{equation}
\label{E:WA4d}
W_A = \int d^4x \sqrt{-g} A_{\mu}\left(c_A j_A^{\mu}+c_m (j_m^{\mu}-\alpha j_{CS}^{\mu})\right)\,,
\end{equation}
with
\begin{align}
\begin{split}
	j_A^{\mu} &= -2\mu\left(B^{\mu}+\frac{\mu}{2}\omega^{\mu}\right), \\
j_m^{\mu} &=-\left(4W^{\mu\nu\rho\sigma}u_{\nu}u_{\rho}\omega_{\sigma}+\frac{1}{3}\left(R+6R_{\nu\rho}u^{\nu}u^{\rho}-6a^2-\frac{9}{2}\omega^2\right)\omega^{\mu} \right), \\
	j_{CS}^{\mu} &=  \epsilon^{\mu\nu\rho\sigma}\left(\Gamma_{\nu\tau}^{\upsilon}\partial_{\rho}\Gamma_{\sigma\upsilon}^{\tau}+\frac{2}{3}\Gamma_{\nu\tau}^{\upsilon}\Gamma_{\rho\phi}^{\tau}\Gamma_{\sigma\upsilon}^{\phi}\right),
\end{split}
\end{align}
where $W^{\mu\nu\rho\sigma}$ is the Weyl tensor. 
The covariant current and stress tensor can be obtained by varying the generating function $W=W_1+W_A$ and adding the appropriate Bardeen-Zumino polynomials~\eqref{E:4dBZterms}, as defined in~\eqref{E:covJT}. The stress tensor and current obtained this way satisfy the anomalous conservation equations~\eqref{E:nonC4d},
\begin{equation}
\begin{split}
\nabla_{\mu} J^{\mu}_{cov} &=  \frac{1}{4}\epsilon^{\kappa\sigma\alpha\beta}\brk{   3 c_{_A}F_{\kappa\sigma}F_{\alpha\beta}
  + c_m R^\nu{}_{\lambda\kappa\sigma} R^\lambda{}_{\nu\alpha\beta} },\\
\nabla_{\nu} T^{\mu\nu}_{cov} &= F^{\mu}_{\phantom{\mu}\nu}J_{cov}^{\nu} 
+2 c_m\  \nabla_\nu \brk{\frac{1}{4}\epsilon^{\kappa\sigma\alpha\beta}F_{\kappa\sigma}R^{\mu\nu}{}_{\alpha\beta} }\,.
\end{split}
\end{equation}
The resulting constitutive relations for the stress tensor and current are presented in appendix \ref{A:4dconstitutive}.

Consider the $T^{\tau r}_{cov}$ component of the stress tensor on a cone times $\mathbb{R}^2$,
\begin{equation}
\label{E:coneR2}
	ds^2 =  dr^2 + r^2 d \tau^2 + dx^2 + dy^2\, ,
\end{equation}
where $\tau \sim \tau+2\pi\delta$, with a magnetic flux given by the gauge potential $A_y =  B x$ with $B$ constant.
We would like to argue that, in this background, the $T^{\tau r}$ component of the stress tensor can only receive contributions from the terms present in~\eqref{E:W14d} and~\eqref{E:WA4d}. To see this, we turn on a metric perturbation $\delta g_{\tau r} = h_{\tau r}(r,x,y)$ and look for terms in the generating function which are linear in the perturbation. Recall that, following section~\ref{S:CS}, we have already accounted for all possible Chern-Simons contributions to the generating function $W$ allowing us to focus our attention on gauge-invariant contributions only. In practice, since all gauge-invariant variables do not depend on the $\mathbb{R}^2$ coordinates $x$ and $y$, it is sufficient to consider $\delta g_{\tau r} = h_{\tau r}(r)$. Indeed, all spatial derivatives of the $\mathbb{R}^2$ coordinates may act only on $h_{\tau r}$ and will therefore lead to a vanishing contribution to the generating function after integrating by parts.
For instance, consider the three derivative contribution to the generating function of the form 
\begin{equation}
	{W} = \int d^4x \sqrt{-g} R_{\mu\nu}u^{\mu}a^{\nu} = i\beta \int dr dx dy  \left( \partial_x^2 + \partial_y^2 \right)\frac{h_{\tau r}}{4r}\,.
\end{equation}
Thus, $\delta {W}/\delta g_{\tau r}(r,x,y) = \delta {W}/\delta g_{\tau r}(r)$. 

Focusing on perturbations of the form $\delta g_{\tau r} =h_{\tau r}(r)$ the Riemann tensor, the vorticity tensor $\Omega_{\mu\nu}$ and the electric field vanish at least to $\mathcal{O}(h^2)$ and
\begin{align}
\begin{split}
\label{E:d4relations1}
	\nabla_{\mu} u_{\nu} &= - u_{\mu}a_{\nu} +\mathcal{O}(h^2) \,,\\
	\nabla_{\mu} T &= -T a_{\mu} \,,\\
	\nabla_{\mu}a_{\nu} &= - u_{\mu}u_{\nu}a^2 -a_{\mu}a_{\nu} + \mathcal{O}(h^2) \,,	 \\
	\nabla_{\mu}B_{\nu} & = B_{\mu}a_{\nu} - a_{\mu}B_{\nu} - a_{\alpha}B^{\alpha} u_{\mu} u_{\nu} + \mathcal{O}(h^2)\,.
\end{split}
\end{align}
It is a straightforward procedure to check that there are no scalars or pseudoscalars which can be constructed out of $u^{\mu}$, $B^{\mu}$ and $a^{\mu}$ which are linear in $h_{\tau r}$.  As was the case in two dimensions, the absence of scalars and pseudo scalars which are linear in $h_{\tau r}$ implies that local non perturbative contributions to $W$ will not affect $T_{cov}^{\tau r}$.

We have argued that in the background given by \eqref{E:coneR2}, the only contribution to $T_{cov}^{\tau r}$ comes from \eqref{E:W14d} and \eqref{E:WA4d}. An explicit computation (see~\ref{E:4d1point}) gives us
\begin{equation}
	T_{cov}^{\tau r} = iB\frac{\tilde{c}_{4d}+8\pi^2\delta^2 c_m}{4\pi^2\delta^2 r^3}.
\end{equation}
We then use the same argument as in the two dimensional case, namely that the translational invariance of the Euclidean vacuum implies that $T_{cov}^{\tau r}$ vanishes in the limit $\delta \to 1$, to obtain
\begin{equation}
\label{E:mixedconstraint}
	\tilde{c}_{4d} = -8 \pi^2 c_m\,.
\end{equation}

\section{Summary and discussion}
\label{S:SandD}

Thermodynamic response coefficients induced by gravitational anomalies pose entirely new types of challenges when compared to response coefficients induced by pure $U(1)$ anomalies. In particular, the gravitational anomaly coefficient $c_g$ in two dimensions and the mixed anomaly coefficient $c_m$ in four dimensions were conjectured to be related to the respective response parameters $\tilde{c}_{2d}$ and $\tilde{c}_{4d}$ via 
\begin{subequations}
\label{eq:GravAnomConj}
\begin{align}
\label{E:GravAnom2d}
	\tilde{c}_{2d}+8\pi^2c_g&=0\\
\label{E:GravAnom4d}
	\tilde{c}_{4d}+8\pi^2c_m&=0\,.
\end{align}
\end{subequations}
These conjectures have a natural generalisation to any even dimension\cite{Loganayagam:2012pz}. They are peculiar in that (a) they mix coefficients of different orders in the derivative expansion and (b) they involve the transcendental number $\pi$. A main aim of this work is to fill up this lacuna in our understanding of gravitational anomaly induced response and to prove \eqref{eq:GravAnomConj}. We will begin our discussion in section \ref{S:Dmethod} by reviewing our derivation of \eqref{eq:GravAnomConj}. Then, in \ref{S:Dlimits} we critically examine various assumptions associated with our technique. We end with a discussion of future prospects in section \ref{S:Dfuture}.

\subsection{Thermodynamics on cones and gravitational anomalies} 
\label{S:Dmethod} 
Consider a two dimensional Euclidean theory on a cone $\mathcal{C}_{\delta}$ of deficit angle $2\pi(\delta-1)$. 
The Euclidean partition function, $W_{\delta}$, of such a theory  can be thought of as the thermal partition function (at a temperature  $T=(2 \pi \delta r)^{-1}$) with the angular direction being the Euclidean time. Since $|\partial_r T|=(2 \pi\delta)T^2$ rather than the required parametric  suppression  for the derivative expansion $|\partial_r T|\ll T^2$, the derivative expansion breaks down on the cone. (This breakdown of the derivative expansion is unrelated to various problems that may originate from the conical singularity at the tip of the  cone, which we will return to in the next subsection.)

A priori, the failure of the derivative expansion means that one cannot directly use the 
methods of \cite{Jensen:2012jh,Banerjee:2012iz} which parametrize thermal partition functions
on arbitrary time-independent backgrounds using the derivative expansion. Despite this,
we argue that one may use the methods in  \cite{Jensen:2012jh,Banerjee:2012iz}
to accurately  compute a specific off-diagonal component  in the one-point function
of the stress tensor $ T^{\tau r}$ of the Euclidean theory.
Further, we argue that $T^{\tau r}$ receives contributions  
from only two terms among the infinite number of terms in the derivative expansion including any local non perturbative contributions to $W$. The two terms on which $T^{\tau r}$ depend are the contribution of the gravitational anomaly proportional to $c_g$ and a certain Chern-Simons term associated with $\tilde{c}_{2d}$.

When there is no deficit angle our generating function reduces to that of the Euclidean vacuum, $W_{Vac}$, which is presumably unique. Assuming that the generating function is continuous at $\delta=1$,
\begin{equation} 
\label{E:limits} 
W_{Vac} =  \lim_{\delta\rightarrow 1}W_{\delta} \,,
\end{equation} 
we demand that $\lim_{\delta \to 1}T^{\tau r} = 0$ resulting in \eqref{E:GravAnom2d}. 
One way of interpreting our computation is to think of $T^{\tau r}$ as a Casimir momentum of the thermal state which is generated in taking  the Euclidean theory from the plane $\mathbb{R}^2$ to the cylinder $\mathbb{S}^1 \times \mathbb{R}$.  
Phrased this way, our argument is conceptually similar to  the argument used to derive the Cardy formula. Indeed, in the framework of conformal field theories we can prove \eqref{E:GravAnom2d} using a formalism similar to that of \cite{Bloete:1986qm,Affleck:1986bv}---see appendix \ref{A:Casimir} for details.
We emphasize that, unlike the Cardy formula, our argument involving the generating function can be applied to non-conformal theories. 
 
Our derivation of \eqref{E:GravAnom2d} (and the Cardy formula argument) satisfactorily circumvent both of the {a priori} obstacles mentioned in the  introduction faced by previous methods: the transcendental 
factor of $8\pi^2$ enters into the  argument geometrically through the radius of the Euclidean circle, and the  breakdown 
of the derivative expansion on the cone allows  the gravitational anomaly coefficient $c_g$ which is second order in the derivative expansion and the coefficient of the Chern-Simons term, $\tilde{c}_{2d}$ which is zero order in the derivative expansion to contribute to  correlation functions at the same order in derivatives generating a ``jump'' in the derivative expansion.
 
The situation in four dimensions is very similar to the two dimensional case. To derive the constraint \eqref{E:GravAnom4d} involving the mixed gravitational anomaly coefficient we place our the theory 
on cone times a plane, $\mathcal{C}_{\delta} \times \mathbb{R}^2$, with a small but a constant magnetic flux  
turned on over $\mathbb{R}^2$. We then consider the off-diagonal component of the stress tensor whose two indices are on the cone. As in the two dimensional case, we can argue that only the terms associated with $\tilde{c}_{4d}$ and $c_m$ 
terms contribute to $T^{\tau r}$. Using the same assumptions as  
those spelt out for the two dimensional case we arrive at \eqref{E:GravAnom4d}. 
 
\subsection{Validity of our results} 
\label{S:Dlimits} 
In this subsection we would like to examine possible pitfalls and the validity of various assumptions underlying 
our derivation of \eqref{eq:GravAnomConj}. The starting point of our analysis was to place 
the field theory under consideration in a background with a conical singularity. To do this in 
practice one must usually regulate the singularity by smoothing out the conical tip 
in some particular way. Following \cite{Fursaev:1995ef} one may consider a class of regulators of the form 
\begin{equation}
\label{E:smoothcone}
	ds_l^2 = r^2 d\tau^2 + F(r/l)dr^2 
\end{equation}
where $F(x)$ is a smooth function of its argument and satisfies $ F(0)= \delta^2$ and $F(\infty)= 1$. For instance, $F = (r^2/\ell^2+\delta^2)/(r^2/\ell^2+1)$ which describes a hyperbolic space satisfies these criteria.

Since the sequence of geometries \eqref{E:smoothcone} are smooth they can not converge uniformly to the cone which is singular. Indeed, one of the main results of \cite{Fursaev:1995ef} is that while various geometric observables (such as arbitrary polynomials of the curvature tensor) depend on the details of the regularization procedure, there are other observables which are independent of such details. We expect this to be true of field theory observables as well. In particular, we expect to find some arbitrariness in the structure of possible localized states near the singularity. (This phenomenon is well-known from  string theory computations involving resolution of orbifold singularities.)



Whether they depend on the regularization parameter or not, the distressing feature of localized states at the tip of the cone is that the Euclidean generating function on the cone, $W_{\delta}$, will not approach the vacuum generating function in the limit $\delta \to 1$, i.e., $W_{\delta}$ is not continuous at $\delta=1$. Since our analysis relied on taking the $\delta\to 1$ limit of the generating function as in \eqref{E:limits}, one might worry that our computation would fail in the presence of localized states at the tip.
However, our analysis focused on computing the $T^{\tau r}$ components of the stress tensor far from the tip of the cone. Therefore, if the equality \eqref{E:limits} is true up to localized expressions at the tip then our computation remains unspoiled. In other words, since we are computing components of the stress tensor away from the tip of the cone, we may require a weaker condition than \eqref{E:limits}. Namely, that the equality in \eqref{E:limits} holds up to terms localized at the tip. Thus, our computation is robust against the addition of any localized state near the singularity.


Despite our computation being robust under the addition of any localized state near the tip, one should still be concerned about the continuity of $W_{\delta}$ at $\delta=1$.  We emphasise that this is not an empty concern. As is known from heat kernel computations
on the cone\cite{Fursaev:1996uz} the heat kernel (and hence the partition function)
of massless higher spin fields (gravitinos and gravitons in this case) 
are not continuous at $\delta=1$. The reason for this discontinuity is the enhanced symmetry on the plane. On the cone we have only one Killing vector, the generator of rotations, while on the plane we have three Killing vectors which generate $ISO(1,1)$, two translations and rotations. Since the graviton and gravitino are susceptible to the geometry the number of zero modes of these fields is discontinuous at $\delta= 1$. 
This suggests, for example, that the relations \eqref{eq:GravAnomConj} will fail for the
gravitational anomaly contribution from chiral gravitino. This was 
in fact known before from explicit computations of $\tilde{c}_{2nd}$ for a free theory of chiral 
gravitinos \cite{loga_WIP}. It is satisfying that our derivation excludes 
those systems where the relations in question are known to break down. 

Instead of placing the theory on a cone $\mathcal{C}_{\delta}$ we could have placed our theory on $\mathbb{R}^{2,*} = \mathbb{R}^2/\{0,0\}$, i.e., the plane with the origin removed. In this case, instead of \eqref{E:limits} we would have had to assume that the origin can be added back without modifying the generating function $W$ away from $\{0,0\}$. In other words, If the generating function $W$ on the punctured plane differs from the generating function $W$ on $\mathbb{R}^2$ only by terms which are localized at the origin then our analysis carries through without any obstruction.

This discussion leads us to conclude that the relations~\eqref{eq:GravAnomConj} hold 
for all local field theories  lacking massless gravitini or gravitons. This is certainly true for free field theories \cite{Fursaev:1996uz} and therefore also for weakly coupled ones.  It would be satisfying to relate our argument to a  mathematical theorem for a specific index which would bypass this problem of continuity entirely.

As we have emphasized several times, the derivative expansion breaks down on the cone. As a result, it is insufficient to use a truncated generating function $W_m$ valid only up to order $m$ in the derivative expansion. Let us restate how this problem has been circumvented. Owing to the fact that in Euclidean space correlators drop of exponentially at small distances as in \eqref{E:screening}, the full Euclidean generating function may be formally expanded in a power series in derivatives around the flat space thermodynamic equilibrium configuration. When resummed, this infinite series together with any non analytic terms at the origin will reconstruct the Euclidean partition function. What we have argued for in sections \ref{S:2d} and \ref{S:4d} is that  there are no terms in this series and no local non analytic terms at the origin which may contribute to the particular component of the stress tensor which we were interested in.

\subsection{Future prospects}
\label{S:Dfuture}
We will now conclude with various thoughts on future work. We begin by noting that the 
parameters $\tilde{c}_{2d}$  and $\tilde{c}_{4d}$ that are the focus of this 
paper are respectively the  coefficients of the 1d  pure and 3d mixed Chern-Simons
terms for the  thermal Kaluza-Klein field. The relations \eqref{eq:GravAnomConj}
essentially relate these Chern-Simons terms to pure or mixed gravitational anomaly
coefficient in the theory with one dimension higher.

A relation between Chern-Simons terms and anomalies is already familiar in the case of $U(1)$ anomalies \cite{AlvarezGaume:1984nf} where an abelian anomaly in $2n$ dimensions is related to a parity anomaly in $2n-1$ dimensions. It would be interesting to see whether a similar argument could be used to directly derive Kaluza-Klein Chern-Simons terms in field theory 
after compactifying on a  warped thermal circle. We refer the reader to~\cite{UW:CVE} for related discussions.


Next, the procedure that we have used to establish \eqref{eq:GravAnomConj}
relied on properties of the partition function on a cone in the limit where the deficit angle
goes to zero. This object has been studied in the literature 
on entanglement entropy (see \cite{Solodukhin:2011gn}   for a review). 
Briefly, given a state in, say, a $3+1$ dimensional field theory, and a plane which  
cuts the three dimensional space into two halves, the entanglement entropy of half the space is the  
Von-Neumann entropy associated with the density matrix obtained after integrating over the of degrees of freedom on one half of the space. An efficient way to compute this entanglement entropy is via the replica trick (which for a field theory  
leads to the so called ``conical singularity'' method) whereby 
one introduces a conical deficit on the interface and studies the resulting 
partition function as a function of the  deficit angle. It would be interesting
to see whether our analysis can be translated into specific statements
about how gravitational anomaly enters the entanglement entropy.

While we have argued for \eqref{eq:GravAnomConj} which are valid in two and four dimensions
it would be interesting to try and prove the higher dimensional generalisation of \eqref{eq:GravAnomConj}
proposed in \cite{Loganayagam:2012pz}. There, a ``replacement rule'' has been proposed where, in the current language, the leading order Chern-Simons coefficients in the generating function are related to Chern-Simons terms and first Pontryagin classes of the anomaly polynomial. It seems more than likely that one may be able to prove the replacement rule by placing a $2n$ dimensional theory on spaces with particular fluxes effectively reducing the problem to the two dimensional one which we have solved. Furthermore, one can hope to generalize the replacement rule of \cite{Loganayagam:2012pz} in such a way that higher order Pontryagin classes in the anomaly polynomial relate to Chern-Simons term in the generating function which are higher order in derivatives.
We leave these and other related problems for future work.

\acknowledgments
We would like to thank Carlos Hoyos, Sachin Jain, Pavel Kovtun, Mukund Rangamani,
Adam Ritz, Piotr Surowka and Cumrun Vafa  for valuable conversations and Dmitri Fursaev for a useful correspondence. We would like to thank Sayantani Bhattacharyya, Sachin Jain, Shiraz Minwalla, Mukund Rangamani, Yaron Oz, and Tarun Sharma for helpful comments on a previous version of this work.
KJ is grateful to the organizers of the program \textbf{Applications of Gauge-Gravity Duality}
at the Technion for their hospitality while this work was in progress. RL and AY would like to thank the organizers of
\textbf{ICTS discussion meeting on String Theory, Bangalore, India} and
the organizers of \textbf{RIKEN BNL Workshop on P- and CP-odd Effects 
in Hot and Dense Matter, Brookhaven, U.S.A.} for their hospitality.
KJ was supported in part by NSERC, Canada. RL is supported by the
Harvard Society of Fellows through a junior fellowship. AY is a Landau 
fellow, supported in part by the Taub foundation as
well as the ISF under grant number $495/11$ and the BSF under grant number
$2014350$.

\begin{appendix}

\section{Casimir energy density and energy flux in two dimensional conformal field theories}
\label{A:Casimir}
Consider a conformal theory with central charges $c_L$ and $c_R$ on an arbitrary manifold with line element
\begin{equation}
\label{E:cline}
	ds^2 = \rho(z,\bar{z}) dz d\bar{z}\,.
\end{equation}	
Using $R = - \square \ln \rho$, the solution to \eqref{E:2dcovCons} is given by
\begin{align}
\begin{split}
\label{E:2dsol}
	T_{cov}^{z\bar{z}} &= - c_w \rho^{-1} \square \ln \rho\,, \\
	T_{cov}^{\bar{z}\bar{z}}& = \frac{2(c_w+2 c_g)}{\rho^4} \left(2 \rho \partial_{z}^2 \rho - 3 (\partial_z \rho)^2 \right) \,, \\
	T_{cov}^{zz} &= \frac{2(c_w-2 c_g)}{\rho^4} \left(2 \rho \partial_{\bar{z}}^2 \rho - 3 (\partial_{\bar{z}} \rho)^2 \right)\,,
\end{split}
\end{align}
where we have required that $T_{cov}^{\mu\nu} = 0$ for $\rho=1$ and defined the linear combinations
\begin{equation}
	c_w + 2 c_g = \frac{c_L}{24\pi}\,,
	\qquad
	c_w - 2 c_g = \frac{c_R}{24\pi}\,,
\end{equation}
in accordance with the conventions of the main text, namely equations \eqref{E:csandcg} and \eqref{eq:cwCFT}.

A cylinder of radius $R$ can be parameterized by the line element $ds^2 = dx^2 + d\theta^2$ with $\theta \sim \theta + 2\pi/R$ which can be obtained from \eqref{E:cline} using 
\begin{equation}
\label{E:cylparam}
	\rho=\frac{R^2}{z\bar{z}}\,,
	\quad
	z=R e^{(x+i\theta)/R}. 
\end{equation}
Inserting \eqref{E:cylparam} into \eqref{E:2dsol} and Wick rotating to Lorentzian signature we obtain
\begin{equation}
\label{E:2dTmn}
	T^{\mu\nu}_{cov} = \begin{pmatrix} 4  \pi^2 c_w & -8 \pi^2 c_g \\
		-8 \pi^2 c_g & 4 \pi^2 c_w 
		\end{pmatrix} T^2
\end{equation}
with $T^{-1}=2\pi R$. Thus, the energy momentum tensor of a conformal theory on a cylinder differs from that of flat space by a Casimir energy and Casimir momentum density proportional to combinations of the left and right moving central charges. Therefore, the energy density and energy flux of a conformal theory are completely specified by the central charges of the theory even though these enter at second order in the derivative expansion. The result \eqref{E:2dTmn} generalizes the results of \cite{Bloete:1986qm,Affleck:1986bv} to theories with a  gravitational anomaly; the off diagonal term of the energy momentum tensor has a universal value on the infinite strip proportional to the difference between the left and right moving central charges.

\section{Constitutive relations associated with anomalies in four dimensional theories.}
\label{A:4dconstitutive}
In this section we list the part of the four dimensional constitutive relations which follow from the anomalous part of the generating functional, $W_A$, which includes the three derivative terms proportional to the mixed anomaly coefficient $c_m$. This is the four dimensional analogue of the two derivative terms derived in~\cite{Valle:2012em} for two dimensional theories with a pure gravitational anomaly.

We will now focus on contributions to the covariant stress tensor and current obtained from $W_A$ in equation \eqref{E:WA4d} by variation, according to the definition~\eqref{E:covJT}. We will denote them by $T^{\mu\nu}_{A\,cov}$, and $J^{\mu}_{A\,cov}$ respectively. We will present these contributions in the following form
\begin{align}
\label{E:4danomConsRel}
\begin{split}
J^{\mu}_{A\,cov} &=\mathcal{N}_A u^{\mu}+\nu_A^{\mu}, \\
T^{\mu\nu}_{A\,cov}& =\mathcal{E}_A u^{\mu}u^{\nu} + \mathcal{P}_A\Delta^{\mu\nu} +u^{\mu}q_A^{\nu}+u^{\nu}q_A^{\mu}+\tau_A^{\mu\nu}, 
\end{split}
\end{align}
with $q_A, \nu_A, \tau_A$ transverse and $\tau_A$ traceleless, and the subscript $A$ denotes the contribution due to the anomalies. We find that the resulting expressions for $\mathcal{E}_A, \,\mathcal{P}_A ,\, \nu^{\mu}_A, \, q^{\mu}_A$, and $\tau^{\mu\nu}_A$ contain a number of Weyl-covariant terms, as well as a number of Weyl-non-covariant terms.\footnote{Because the gauge and mixed gravitational anomalies may be
non-vanishing in a CFT, we find it interesting that there are
non-Weyl-covariant terms in~\eqref{E:4danomConsRel}. We leave a
comprehensive understanding of these terms and their relation to the
Weyl anomaly for future work.} For this reason, we have found it useful to summarize  a number of independent Weyl-covariant tensor structures in Tables~\ref{T:4dWtensors1} and~\ref{T:4dWtensors2}. There are also the Weyl-covariant zeroth order quantities $T,\mu$ and $u^{\mu}$, as well as the first order pseudovectors $\omega^{\mu}$ and $B^{\mu}$ defined through
\begin{equation}
\omega^{\mu} = \epsilon^{\mu\nu\rho\sigma}u_{\nu}\nabla_{\rho}u_{\sigma}, \qquad B^{\mu}=\frac{1}{2}\epsilon^{\mu\nu\rho\sigma}u_{\nu}F_{\rho\sigma}\,,
\end{equation} 
and the antisymmetric tensor $\Omega^{\mu\nu}$ defined in the caption of Table~\ref{T:4dWtensors1}.
\begin{table}[tc]
\begin{center}
\begin{tabular}{|c|c|c|}
\hline & 1 & 2 \\
\hline 2nd order scalar ($s_{2,i}$ ) & $R+6 R_{\mu\nu}u^{\mu}u^{\nu}-6a^2$ & $u_{\mu}\nabla_{\nu}F^{\mu\nu}$ \\
2nd order vector ($v_{2,i}$) & $\Delta^{\mu\nu}R_{\nu\rho}u^{\rho}-2\Omega^{\mu\nu}a_{\nu}$ & $\Delta^{\mu\rho}\nabla^{\nu}F_{\rho\nu}$ \\
 2nd order tensor ($t_{2}$) &  $W^{\mu\rho\sigma\nu}u_{\rho}u_{\sigma}$ & \\
\hline
\end{tabular}
\caption{\label{T:4dWtensors1} Some independent Weyl-covariant second-order tensors, where we have defined antisymmetric tensor $\Omega^{\mu\nu}=\Delta^{\mu\rho}\Delta^{\nu\sigma}(\nabla_{\rho}u_{\sigma}-\nabla_{\sigma}u_{\rho})/2$. The other independent second-order tensors are products of first-order tensors.}
\end{center}
\end{table}
\begin{table}[tc]
\begin{center}
\begin{tabular}{|c|c|}
\hline & \\
\hline 3rd order pseudovector ($\tilde{v}_{3}$) & $\epsilon^{\mu\nu\rho\sigma}u_{\nu}(\nabla_{\rho}R_{\sigma\alpha})u^{\alpha} + a^{(\nu}\Delta^{\rho )\mu}\nabla_{\nu}\omega_{\rho}+\omega^{\mu} a^2$ \\
3rd order pseudotensor ($\tilde{t}_{3}$) & $\Delta^{\alpha <\mu}\epsilon^{\nu> \rho\sigma\tau}u_{\rho}\left( \nabla_{\sigma}R_{\tau \alpha}+2a_{\sigma}t_{2\,\tau\alpha} \right) $ \\
\hline
\end{tabular}
\caption{\label{T:4dWtensors2} Some Weyl-covariant third-order pseudotensors.}
\end{center}
\end{table}

In terms of these Weyl covariant quantities, we find that the scalars $\mathcal{N}_A$, $\mathcal{E}_A$, and $\mathcal{P}_A$ are given by
\begin{subequations}
\label{E:4d1point}
\begin{equation}
\label{E:4dscalar}
\mathcal{N}_A = 0, \qquad \mathcal{E}_A = 3\mathcal{P}_A = 2c_m \omega_{\mu}(\mu v_{2,1}^{\mu}-v_{2,2}^{\mu})+4c_mB_{\mu}(v_{2,1}^{\mu}-\Omega^{\mu\nu}a_{\nu}).
\end{equation}
The vectors $\nu^{\mu}_A$ and $q^{\mu}_A$ are given by
\begin{align}
	\nu^{\mu}_A =& -6 c_A \mu B^{\mu} - 3 c_A \mu^2 \omega^{\mu}-4c_m t_2^{\mu\nu}\omega_{\nu} - \frac{4}{3}c_m\left( s_{2,1}-\frac{9}{2}\omega^2\right)\omega^{\mu},\\
\nonumber	q^{\mu}_A =& - 3 c_A \mu^2 B^{\mu}
			- 2 c_A \mu^3\omega^{\mu} 
			-2c_m\left(\Phi^{\mu\nu}B_{\nu}+\left(\frac{\omega^2}{4}-a^2\right)B^{\mu} \right) \\
\nonumber			& -2\mu c_m \left( \tilde{v}_{3}^{\mu}+t_{2}^{\mu\nu}\omega_{\nu}+\frac{1}{3}\left(s_{2,1}-\frac{3}{2}\omega^2\right)\omega^{\mu} \right) + c_m \left( \Delta^{\mu\nu}E^{\rho}\nabla_{\rho}\omega_{\nu} + 2 \omega^{\mu} E_{\nu} a^{\nu} - E^{\mu} a_{\nu}\omega^{\nu}\right),
\end{align}
where we have defined the transverse tensor $\Phi_{\mu\nu}=R_{\mu\rho\nu\sigma}u^{\rho}u^{\sigma}$. Finally the tensor $\tau^{\mu\nu}_A$ is
\begin{align}
\tau^{\mu\nu}_A =
	& 4\mu c_m \left( \tilde{t}_{3}^{\mu\nu} - W^{\rho <\mu\nu> \sigma}u_{\rho}\omega_{\sigma} 
	+ 2\omega^{<\mu}v_{2,1}^{\nu>}\right)+2c_m\omega^{<\mu}v_{2,2}^{\nu>}
	+4c_m \omega^{<\mu}\Omega^{\nu>\rho}E_{\rho} \\
	\nonumber & + 4c_m B^{<\mu}v_{2,1}^{\nu>} + 2c_m \epsilon^{\rho\sigma\tau<\mu}u_{\rho}E_{\sigma}(2t_{2\,\tau}^{\phantom{2\,}\nu>}-R_{\tau\alpha}\Delta^{\nu> \alpha}) \\
	\nonumber &- 2 c_m \nabla^{<\mu}F^{\nu>\rho}\omega_{\rho} 
	+ 4c_m a^{<\mu}\epsilon^{\nu>\rho\sigma\tau}u_{\rho}B_{\sigma}\omega_{\tau},
\end{align}
\end{subequations}
where the angular brackets denote the transverse traceless projection
\begin{equation}
V^{<\mu\nu>} = \Delta^{\mu\rho}\Delta^{\nu\sigma}V_{(\rho\sigma)}-\frac{1}{3}\Delta^{\mu\nu}\Delta_{\rho\sigma}V^{\rho\sigma}.
\end{equation}
As expected from the arguments given in section~\ref{S:intro}, all of the expressions in~\eqref{E:4d1point} proportional to $c_m$ have three derivatives and so do not relate the mixed anomaly coefficient $c_m$ to the first derivative term $\tilde{c}_{4d}$ listed in~\eqref{E:4dconstRel}.

\end{appendix} 

\bibliographystyle{JHEP}
\bibliography{Mixed}
\end{document}